\documentclass[iop]{emulateapj}

\usepackage{mathtools}

\begin{document}


\newcommand{\vdag}{(v)^\dagger}

\newcommand{\lc}{light curve~}
\newcommand{\lct}{light curve}
\newcommand{\lcs}{light curves~}
\newcommand{\lcst}{light curves}
\newcommand{\Lc}{Light curve~}
\newcommand{\Lcs}{Light curves}
\newcommand{\avg}[1]{\ensuremath{\langle #1\rangle}}
\newcommand{\oot}{out-of-transit~}
\newcommand{\OOT}{Out-of-Transit~}

\newcommand{\fov}[2]{\ensuremath{ \rm #1.\!^\prime #2 \times \rm #1.\!^\prime #2}}
\newcommand{\fova}[4]{\ensuremath{ \rm #1.\!^\prime #2 \times \rm #3.\!^\prime #4}}
\newcommand{\pxs}[2]{\ensuremath{#1.\!^{\prime\prime}#2\ \rm{pixel}^{-1}}}
\newcommand{\bin}[1]{\ensuremath{ \rm #1 \times \rm #1} \rm{pixel}}

\newcommand{\rsun}{\ensuremath{R_\sun}}
\newcommand{\msun}{\ensuremath{M_\sun}}
\newcommand{\lsun}{\ensuremath{L_\sun}}
\newcommand{\teffsun}{\ensuremath{T_{eff,\sun}}}
\newcommand{\rhosun}{\ensuremath{\rho_\sun}}

\newcommand{\rstar}{\ensuremath{R_s}}
\newcommand{\mstar}{\ensuremath{M_s}}
\newcommand{\lstar}{\ensuremath{L_s}}
\newcommand{\astar}{\ensuremath{a_s}}
\newcommand{\loglstar}{\ensuremath{\log{L_s}}}
\newcommand{\teffstar}{\ensuremath{T_{eff,s}}}
\newcommand{\rhostar}{\ensuremath{\rho_s}}

\newcommand{\rearth}{\ensuremath{R_\earth}}
\newcommand{\mearth}{\ensuremath{M_\earth}}
\newcommand{\learth}{\ensuremath{L_\earth}}
\newcommand{\teffearth}{\ensuremath{T_{eff,\earth}}}
\newcommand{\rhoearth}{\ensuremath{\rho_\earth}}

\newcommand{\rpl}{\ensuremath{R_{p}}}
\newcommand{\mpl}{\ensuremath{M_{p}}}
\newcommand{\lpl}{\ensuremath{L_{p}}}
\newcommand{\teffpl}{\ensuremath{T_{eff,{p}}}}
\newcommand{\rhopl}{\ensuremath{\rho_{p}}}
\newcommand{\ipl}{\ensuremath{i_{p}}}
\newcommand{\epl}{\ensuremath{e_{p}}}
\newcommand{\gpl}{\ensuremath{g_{p}}}
\newcommand{\fpl}{\ensuremath{f_{p}}}

\newcommand{\rjup}{\ensuremath{R_{\rm J}}}
\newcommand{\mjup}{\ensuremath{M_{\rm J}}}
\newcommand{\ljup}{\ensuremath{L_{\rm J}}}
\newcommand{\teffjup}{\ensuremath{T_{eff,{\rm J}}}}
\newcommand{\rhojup}{\ensuremath{\rho_{\rm J}}}
\newcommand{\gjup}{\ensuremath{\g_{\rm J}}}

\newcommand{\rjuplong}{\ensuremath{R_{\rm Jup}}}
\newcommand{\mjuplong}{\ensuremath{M_{\rm Jup}}}
\newcommand{\ljuplong}{\ensuremath{L_{\rm Jup}}}
\newcommand{\teffjuplong}{\ensuremath{T_{eff,{\rm Jup}}}}
\newcommand{\rhojuplong}{\ensuremath{\rho_{\rm Jup}}}
\newcommand{\gjuplong}{\ensuremath{\g_{\rm Jup}}}

\newcommand{\teff}{\ensuremath{T_{\rm eff}}}
\newcommand{\teq}{\ensuremath{T_{\rm eq}}}
\newcommand{\logg}{\ensuremath{\log g}}

\newcommand{\vsini}{\ensuremath{V\sin(I)}}

\newcommand{\kms}{km~s$^{-1}$}
\newcommand{\ms}{m~s$^{-1}$}

\newcommand{\msini}{\ensuremath{m \sin i}}
\newcommand{\mplsini}{\ensuremath{\mpl\sin i}}
\newcommand{\mtsini}{\ensuremath{M_{2}\sin i}}

\newcommand{\ag}{\ensuremath{A_{\rm g}}}
\newcommand{\ab}{\ensuremath{A_{\rm B}}}
\newcommand{\eps}{\ensuremath{\varepsilon}}
\newcommand{\tn}{\ensuremath{T_{\rm N}}}
\newcommand{\td}{\ensuremath{T_{\rm D}}}
\newcommand{\tdeff}{\ensuremath{T_{\rm D, eff}}}

\newcommand{\tmid}{\ensuremath{T_{\rm mid}}}
\newcommand{\tocc}{\ensuremath{T_{\rm occ}}}
\newcommand{\ttr}{\ensuremath{T_{\rm tr}}}

\newcommand{\dkep}{\ensuremath{D_{\ik}}}

\newcommand{\ik}{{\it Kepler}}

\newcommand{\is}{{\it Spitzer}}

\newcommand{\ks}{$K_{s}$}
\newcommand{\kp}{$K_{p}$}

\newcommand{\mic}{\ensuremath {\rm \mu m}}

\newcommand{\sig}[1]{\ensuremath{#1\sigma}}

\newcommand{\pref}[1]{p.~\pageref{#1}}
\newcommand{\figr}[1]{Fig.~\ref{fig:#1}}
\newcommand{\figsr}[2]{Figs.~\ref{fig:#1}~\&~\ref{fig:#2}}
\newcommand{\secr}[1]{Sec.~\ref{sec:#1}}
\newcommand{\appr}[1]{Appendix~\ref{app:#1}}
\newcommand{\secsr}[2]{Secs.~\ref{sec:#1}~\&~\ref{sec:#2}} 
\newcommand{\eqr}[1]{Eq.~\ref{eq:#1}}
\newcommand{\eqsr}[2]{Eqs.~\ref{eq:#1}~and~\ref{eq:#2}}
\newcommand{\tabsr}[1]{Tab.~\ref{tab:#1}}
\newcommand{\tabr}[1]{\mbox{Table~\ref{tab:#1}}}
\newcommand{\figrp}[1]{Fig.~\ref{fig:#1} on \pref{fig:#1}}
\newcommand{\secrp}[1]{\S\ref{sec:#1} on \pref{sec:#1}}
\newcommand{\eqrp}[1]{Eq.~\ref{eq:#1} on \pref{eq:#1}}
\newcommand{\tabrp}[1]{Tab.~\ref{tab:#1} on \pref{tab:#1}}

\newcommand{\Abea}{\ensuremath{A_{\rm beam}}}
\newcommand{\albea}{\ensuremath{\alpha_{\rm beam}}}
\newcommand{\albeanu}{\ensuremath{\alpha_{\rm beam, \nu}}}
\newcommand{\Aell}{\ensuremath{A_{\rm ellip}}}
\newcommand{\alell}{\ensuremath{\alpha_{\rm ellip}}}
\newcommand{\Aref}{\ensuremath{A_{\rm refl}}}
\newcommand{\alref}{\ensuremath{\alpha_{\rm refl}}}


\def\aj{AJ}%
\def\actaa{Acta Astron.}%
\def\araa{ARA\&A}%
\def\apj{ApJ}%
\def\apjl{ApJ}%
\def\apjs{ApJS}%
\def\ao{Appl.~Opt.}%
\def\apss{Ap\&SS}%
\def\aap{A\&A}%
\def\aapr{A\&A~Rev.}%
\def\aaps{A\&AS}%
\def\azh{AZh}%
\def\baas{BAAS}%
\def\bac{Bull. astr. Inst. Czechosl.}%
\def\caa{Chinese Astron. Astrophys.}%
\def\cjaa{Chinese J. Astron. Astrophys.}%
\def\icarus{Icarus}%
\def\jcap{J. Cosmology Astropart. Phys.}%
\def\jrasc{JRASC}%
\def\mnras{MNRAS}%
\def\memras{MmRAS}%
\def\na{New A}%
\def\nar{New A Rev.}%
\def\pasa{PASA}%
\def\pra{Phys.~Rev.~A}%
\def\prb{Phys.~Rev.~B}%
\def\prc{Phys.~Rev.~C}%
\def\prd{Phys.~Rev.~D}%
\def\pre{Phys.~Rev.~E}%
\def\prl{Phys.~Rev.~Lett.}%
\def\pasp{PASP}%
\def\pasj{PASJ}%
\def\qjras{QJRAS}%
\def\rmxaa{Rev. Mexicana Astron. Astrofis.}%
\def\skytel{S\&T}%
\def\solphys{Sol.~Phys.}%
\def\sovast{Soviet~Ast.}%
\def\ssr{Space~Sci.~Rev.}%
\def\zap{ZAp}%
\def\nat{Nature}%
\def\iaucirc{IAU~Circ.}%
\def\aplett{Astrophys.~Lett.}%
\def\apspr{Astrophys.~Space~Phys.~Res.}%
\def\bain{Bull.~Astron.~Inst.~Netherlands}%
\def\fcp{Fund.~Cosmic~Phys.}%
\def\gca{Geochim.~Cosmochim.~Acta}%
\def\grl{Geophys.~Res.~Lett.}%
\def\jcp{J.~Chem.~Phys.}%
\def\jgr{J.~Geophys.~Res.}%
\def\jqsrt{J.~Quant.~Spec.~Radiat.~Transf.}%
\def\memsai{Mem.~Soc.~Astron.~Italiana}%
\def\nphysa{Nucl.~Phys.~A}%
\def\physrep{Phys.~Rep.}%
\def\physscr{Phys.~Scr}%
\def\planss{Planet.~Space~Sci.}%
\def\procspie{Proc.~SPIE}%
\let\astap=\aap
\let\apjlett=\apjl
\let\apjsupp=\apjs
\let\applopt=\ao


\slugcomment{PASP Invited Review}

\title{The astrophysics of visible-light orbital phase curves in the space age}
\shorttitle{Phase curves in the space age}

\author{Avi Shporer}
\affil{Division of Geological and Planetary Sciences, California Institute of Technology, Pasadena, CA 91125, USA}
\email{shporer@gps.caltech.edu}
\shortauthors{Shporer}

\begin{abstract}

The field of visible-light continuous time series photometry is now at its golden age, manifested by the continuum of past (CoRoT, \ik), present (K2), and future (TESS, PLATO) space-based surveys delivering high precision data with a long baseline for a large number of stars. The availability of the high quality data has enabled astrophysical studies not possible before, including for example detailed asteroseismic investigations and the study of the exoplanet census including small planets. This has also allowed to study the minute photometric variability following the orbital motion in stellar binaries and star-planet systems which is the subject of this review. We focus on systems with a main sequence primary and a low-mass secondary, from a small star to a massive planet. The orbital modulations are induced by a combination of gravitational and atmospheric processes, including the beaming effect, tidal ellipsoidal distortion, reflected light, and thermal emission. Therefore, the phase curve shape contains information about the companion's mass and atmospheric characteristics, making phase curves a useful astrophysical tool. For example, phase curves can be used to detect and measure the mass of short-period low-mass companions orbiting hot fast-rotating stars, out of reach of other detection methods. Another interesting application of phase curves is using the orbital phase modulations to look for non-transiting systems, which comprise the majority of stellar binary and star-planet systems. We discuss the science done with phase curves, the first results obtained so far, and the current difficulties and open questions related to this young and evolving subfield.

\end{abstract}


\section{Introduction}
\label{sec:intro}

Observational astrophysics is often led by technological development. The availability of large CCDs at the end of the 20th century combined with the development of automated observatories has enabled wide field photometric surveys capable of simultaneously monitoring a large number of stars. This has revolutionized the field of variable stars and the wide range of astrophysics variable stars make possible \citep{paczynski97, paczynski00}. 

The first such surveys were initiated in the early 1990's. While some were designed and operated to obtain a census of variable stars (e.g.~ASAS, \citealt{pojmanski97}), others pursued a specific type of variability, such as microlensing (OGLE, \citealt{udalski92}; MACHO, \citealt{alcock00}; EROS, \citealt{aubourg93}), gamma ray burst optical afterglow (ROTSE, \citealt{akerlof99}), and Supernovae (BAIT, \citealt{richmond93}). 

In the late 1990's the discovery of short period gas giant planets orbiting Sun-like stars \citep[e.g.,][]{mayor95, butler97} motivated constructing surveys designed to detect the photometric transit signal as a planet transits across the disk of its host star (e.g., HAT, \citealt{bakos04}; KELT, \citealt{pepper07}; TrES, \citealt{alonso04}; Vulcan, \citealt{borucki01}; WASP, \citealt{pollacco06}; XO, \citealt{mccullough05}).

The success of the ground-based surveys has led to carrying out similar surveys using dedicated space-based telescopes, capable of reaching far greater photometric precision while conducting temporally uninterrupted monitoring. The high-quality data delivered by space-based surveys such as CoRoT \citep{auvergne09} and \ik\ \citep{borucki11, borucki16} opened the door to scientific investigations that were not possible before using ground-based data, showing again that observational astronomy is often dominated by the available instrumentation. This is exemplified by the discovery of small planets \citep[e.g.,][]{leger09, batalha11, borucki11} and the study of their occurrence \citep[e.g.,][]{petigura13, burke15, winn15}, and detailed asteroseismic studies \citep[e.g.,][]{gilliland10, chaplin11, chaplin13}.

Another scientific investigation that was enabled by space-based data is that of orbital phase curves --- the photometric variability induced by the orbital motion in a two-body system. While phase curves were, and are, observed from the ground as well, such observations are done for systems that typically include compact objects on short periods \citep[e.g.,][]{shakura87, maxted00, gelino01, geier07, shporer07, shporer10}, where the photometric amplitudes are orders of magnitude larger than in systems including regular stars and low-mass objects (from low-mass stars, to brown dwarfs, to planets).

The detection of phase curve variability in two-body systems containing regular stars and low-mass objects down to the planet mass was discussed already before space-based data were available \citep{jenkins03, loeb03, zucker07, pfahl08}. In this review we discuss the first results from the study of phase curves obtained by space-based surveys. We focus on visible-light (optical) data, the wavelength range where these surveys operate, and on systems with a Sun-like main sequence primary star and low-mass secondary down to the planet mass. 

This paper is organized as follows. In \secr{processes} we describe the physical processes inducing variability along the orbital phase. In \secr{astro} we describe several ways of using orbital phase curves as an astrophysical tool while discussing the first results and current limitations of each application (\secr{hotstars} -- \secr{atm}). We also show, in \secr{massdiscrep}, how the recent studies have revealed gaps in the understanding of the phase curve shape and while this has already led to improved understanding some gaps still remain. In \secr{future} we discuss future prospects and we conclude with a summary in \secr{sum}.

\section{The physical processes inducing photometric variability correlated with the orbit}
\label{sec:processes}

The physical process inducing variability along the orbit can be broadly divided into two classes: gravitational processes and atmospheric processes. The gravitational processes include two processes, the beaming effect, described in \secr{beam}, and tidal ellipsoidal distortion, described in \secr{ellip}. The sinusoidal orbital modulation induced by each of these two processes is independent of the other hence they produce two phase modulation components. The atmospheric processes also include two processes, reflected light and thermal emission, described in \secr{ref}. However, we expect these two processes to show the same phase modulation shape hence they both contribute to a single atmospheric phase modulation component. Therefore we have a total of three phase modulations components, described in the subsections below and presented schematically in \figr{scheme}. The analytic description of the beaming effect amplitude is relatively precise, compared to the ellipsoidal distortion and atmospheric phase components whose analytic description involves some approximations and assumptions.

\subsection{The beaming effect (Doppler boosting)}
\label{sec:beam}

The photometric beaming effect, also called Doppler boosting, causes the observed flux of a celestial object to be dependent on the relative radial velocity (RV) between the object and the observer. It is composed of three processes: Doppler shift, time dilation, and light abberation \citep[e.g.,][]{vankerkwijk10, bloemen11, prsa16}. 

Assuming that $v_r \ll c$, where $v_r$ is the object's RV and $c$ is the speed of light, the transformation between the emitted flux $F_{0,\nu}$ at frequency $\nu$ to the observed flux $F_{\nu}$ is:
\begin{equation}
\label{eq:trans}
F_{\nu} = F_{0,\nu} (1 + \albeanu 4 \frac{v_r}{c})\ ,
\end{equation}
where \albeanu\ is an order of unity coefficient discussed in more details below.

\begin{figure}[h]
\hspace{-10mm}
\includegraphics[width=10.5cm]{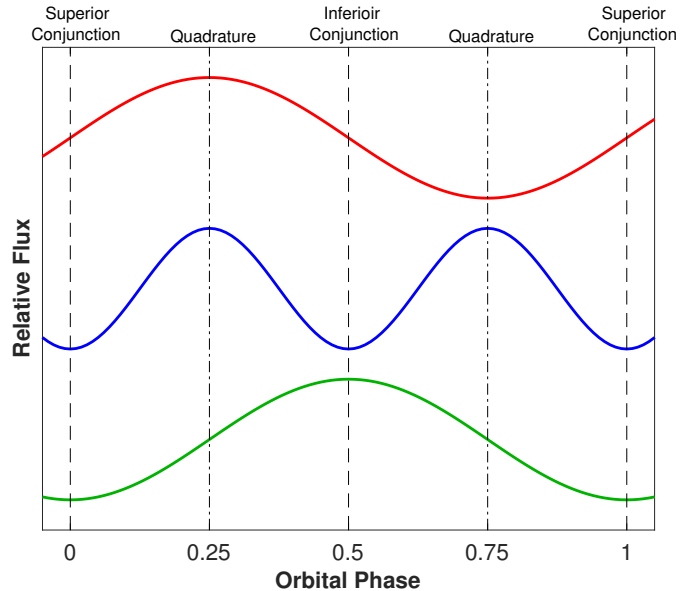}
\caption{Schematic light curve shape of the three orbital phase components, including beaming (top, red, see \secr{beam}), ellipsoidal (middle, blue, see \secr{ellip}), and atmospheric (bottom, green, see \secr{ref}). As marked in the figure, phases 0.25 and 0.75 are quadrature, phase 0.0 (and 1.0) is inferior conjunction, and phase 0.5 is superior conjunction. Conjunction phases are where eclipses would occur if the two-body system is eclipsing. Primary eclipse, or transit, would be at inferior conjunction and secondary eclipse would occur at superior conjunction \citep[e.g.][]{winn10}. These light curve shapes are simplistic and represent low order approximations. They are plotted with the same amplitude and shifted vertically for clarity. A circular orbit is assumed and the atmospheric component includes both reflected light and thermal emission.
}
\label{fig:scheme}
\end{figure}

Following the above, a variation in the RV leads to a variation in the observed flux. In binary systems, where the RV changes periodically as the two objects orbit the common center of mass, this results in a corresponding photometric modulation along the orbit. The shape of the orbital flux modulation induced by the beaming effect is identical to that of the RV curve, albeit with opposite signs since the RV is defined to be {\it positive} when the object is moving {\it away} from the observer, and that motion causes a {\it decrease} in measured flux due to the beaming effect. The photometric modulation induced by the beaming effect reaches extrema at quadrature phases. It is at maximum during the quadrature phase where the target is moving towards the observer, and at minimum at the quadrature where it is moving away, as illustrated schematically in \figr{scheme} by the red curve.

Following \eqr{trans}, a simple way to express the fractional photometric amplitude of the beaming effect induced by orbital motion in a binary system, \Abea, is
\begin{equation}
\label{eq:abeam1}
\Abea = \albea 4 \frac{K_{RV}}{c},
\end{equation}
where the photometric amplitude is linear in the orbital RV amplitude $K_{RV}$ of the same object, and \albea\ is the wavelength integrated \albeanu. Another way to express \Abea\ is by using the system's physical parameters:
\begin{equation}
\begin{multlined}
\label{eq:abeam2}
\Abea = 0.0028\ \albea \left(\frac{P}{\rm day}\right)^{-1/3}\\
\times\left(\frac{M_1+M_2}{\msun}\right)^{-2/3} \left(\frac{M_2 \sin i}{\msun}\right) ,
\end{multlined}
\end{equation}
which is the photometric beaming amplitude in relative flux of an object with mass $M_1$ as it orbits a companion with mass $M_2$ at an orbital period $P$, and $i$ is the orbital plane inclination angle. In case where $M_2 \ll M_1$ \eqr{abeam2} can be approximated as:
\begin{equation}
\begin{multlined}
\label{eq:abeam3}
\Abea \approx 2.7\ \albea \left(\frac{P}{\rm day}\right)^{-1/3}\\
\times\left(\frac{M_1}{\msun}\right)^{-2/3} \left(\frac{M_2 \sin i}{\mjup}\right)\ \rm ppm\ ,
\end{multlined}
\end{equation}
where \mjup\ is Jupiter mass and ppm stands for part per million or 10$^{-6}$.

We focus now on the \albea\ coefficient. In bolometric light $\albea \equiv 1$ \cite[e.g.,][]{loeb03, zucker07}\footnote{Some authors define \albea\ to have a bolometric value of 4 \citep[e.g.,][]{vankerkwijk10, bloemen11} thus eliminating the prefactor of 4 from \eqsr{trans}{abeam1}.}, while in a finite wavelength range \albea\ may deviate from unity. As the observed object moves in its orbit the emitted photons are periodically blue shifted and red shifted. Therefore the amount of observed photons per unit wavelength range modulates. This Doppler shift may lead to a modulation in the total amount of observed photons, depending on the shape of the transmission curve and the observed object's spectrum. Whether this increases or decreases the overall beaming amplitude is quantified by the deviation of \albea\ from unity.

\begin{figure}[h]
\centering
\includegraphics[scale=0.47]{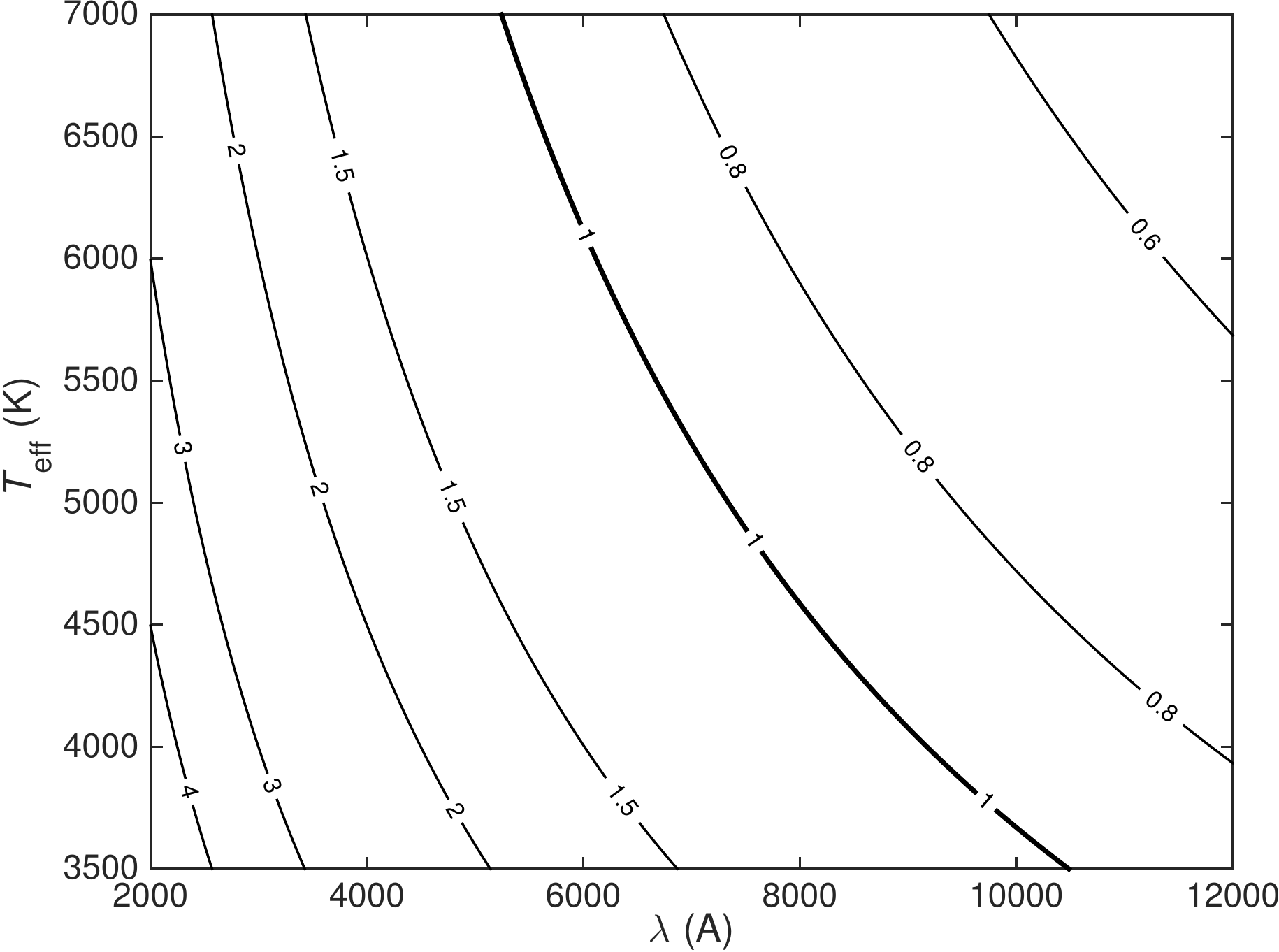}
\caption{
\label{fig:albeam}
Contour plot showing the value of the beaming coefficient \albeanu\ as a function of wavelength $\lambda$ ($x$-axis) and effective temperature \teff\ ($y$-axis), while assuming a target with a black body spectrum (See Eqs.~5 and 6). The \albeanu\ values are labeled along the contours. For a target with a given \teff\ observed with a given transmission curve \albea\ is derived by integrating over the wavelength axis while weighing the integral by the transmission curve. The figure shows that \albeanu\ increases with decreasing wavelength and decreasing temperature.
}
\end{figure}

The \albea\ coefficient is closely related to the spectral flux index. At a specific observed frequency it is \citep[e.g.,][]{loeb03, zucker07, vankerkwijk10}:
\begin{equation}
\albeanu = \frac{1}{4} \left( 3 - \frac{d \log F_{\nu}}{d \log \nu} \right) \ .
\end{equation}
Numerical estimates of \albeanu\ can be done by using an observed spectrum of the target spanning the photometric band, or, by using a synthetic spectrum. Assuming (or approximating) the object is a blackbody allows to determine \albeanu\ analytically:
\begin{equation}
\label{eq:albeanu}
\albeanu = \frac{1}{4}\frac{x e^x}{e^x-1}\ , \ \ \ \ \ \ \ {\rm where}\ \ \  x \equiv \frac{h \nu}{k\teff}\ ,
\end{equation}
where \teff\ is the object's effective temperature, $h$ is Planck's constant, and $k$ is the Boltzmann constant. \albea\ is then calculated by integrating \albeanu\ across the target's spectrum weighted by the transmission curve. We show in \figr{albeam} the dependence of \albeanu\ on wavelength and target's effective temperature. The figure shows how \albeanu\ increases for cooler stars and at shorter wavelength.

\subsubsection{The Photometric RM Effect (Rotational Doppler Beaming)}
\label{sec:prm}

The beaming effect impacts the light curve shape also within eclipses. That effect can be perceived as the photometric analog of the Rossiter-McLaughlin effect \citep[RM;][]{gaudi07}, resulting from applying the beaming effect to the rotating stellar surface of the star being eclipsed.

As the eclipsed star is rotating, the beaming effect causes the stellar surface brightness to depend on the surface velocity component towards the observer. This in turn makes the local surface brightness at a given position dependent on the distance from the stellar rotation axis, and whether that position is on the hemisphere rotating towards or away from the observer.

Therefore, as the eclipsing object is moving (or transiting) across the disk of the star being eclipsed it generates an anomalous photometric signal, analogous to the spectroscopic signal of the RM effect. This is referred to as the Photometric RM effect \citep[PRM;][]{shporer12} or rotational Doppler beaming \citep{groot12}. Similarly to the spectroscopic RM signal, the PRM signal holds information about the obliquity of the eclipsed star. This effect was first discussed in the context of binaries consisting of compact objects \citep{hills74}, and then in the context of binaries composed of regular stars and high quality photometric data \citep{shporer12, groot12}.

As mentioned above, the PRM effect causes a stellar surface brightness distribution that is {\it anti-symmetric} about the stellar spin axis. This is in addition to the more well known surface brightness distribution due to limb darkening, which has a radial symmetry about the center of the stellar disk. In addition, for fast rotating stars, with equatorial rotation velocity at the level of a few 10 \kms\ and beyond (as measured for example for hot early-type stars) the surface brightness distribution is impacted also by gravity darkening following the stars' oblate shape. The latter has an axial symmetry and hence can also be used to measure the obliquity of stars in eclipsing systems \citep{barnes09, barnes11, ahlers15, masuda15} although with some degeneracies. Since the PRM-induced surface brightness distribution is axially anti-symmetric and that of gravity darkening is axially symmetric, measuring both can lift some of the degeneracies involved with measuring only one of them \citep{barnes11}. However, at the time of this writing no measurement of the PRM effect was reported in the literature.

\subsection{Tidal ellipsoidal distortion}
\label{sec:ellip}

In binary systems where the two stars are sufficiently close to each other the tidal force that one component induces on the other is strong enough to tidally distort the latter in a manner that generates photometric orbital modulations. The distorted star becomes slightly ellipsoidal in shape, increasing the distance between the stellar surface and its center along the line connecting the centers of the two stars, and decreasing that distance on the plane orthogonal to that line. Regions on the stellar surface that are farther away from the center experience gravity darkening, where the decreased gravity leads to decreased surface temperature and hence decreased surface brightness compared to regions on the stellar surface with a shorter distance to its center. This surface brightness distribution causes a photometric modulation along the orbit that reaches maximum at the two quadratures, when the observer views the elongated profile of the star(s) and reaches minimum at conjunctions when the observers views the narrow profile of the star(s), as shown schematically by the blue curve in \figr{scheme}. Therefore, the photometric modulation is at the first harmonic of the orbital period. The amplitude of the ellipsoidal photometric modulation is approximated by the following analytic expression:
\begin{equation}
\begin{multlined}
\label{eq:ellip}
\Aell \simeq \alell \frac{M_2 \sin i}{M_1} \left( \frac{R_1}{a}\right)^{3}  \sin i 
= 13\ \alell \sin i \\
\times \left( \frac{R_1}{\rsun}\right)^{3} \left( \frac{M_1}{\msun}\right)^{-2} \left( \frac{P}{\rm day}\right)^{-2}   \left( \frac{M_2 \sin i}{\mjup}\right)   \rm ppm\ .
\end{multlined}
\end{equation}
The analytic description equals the tidal force star 2 (companion) raises on star 1 (primary), divided by the surface gravity of star 1, and the \alell\ coefficient accounts for the stellar limb darkening and gravity darkening:
\begin{equation}
\label{eq:alell}
\alell = 0.15 \frac{(15+u)(1+g)}{3-u} ,
\end{equation}
where $g$ is the gravity darkening coefficient and $u$ the linear limb darkening coefficient. For a tabulation of these coefficients values see \cite{claret11}.

\eqr{ellip} is a simplistic description, ignoring the impact of higher order processes such as a phase lag of the tidal bulge, non-synchronous rotation (when the stellar rotation is faster than the orbit), misalignment between the orbital angular momentum and stellar spin, and non-circular orbits. 

More accurate analytic descriptions of \Aell\ can be found in, e.g., the studies of \citet{morris85} and \citet[][see also \citealt{kopal59} Chapter IV]{morris93} showing that \eqr{ellip} is the lowest order term in a Fourier series including higher orders of $R_1/a$. For photometric data with a precision of about 1\% \eqr{ellip} gives a sufficiently accurate description. For higher quality data, obtained typically by space-based photometric monitoring surveys, the data occasionally show deviations from the simplistic description of \eqr{ellip} and require additional higher order terms including additional harmonics \citep[e.g.,][]{jackson12, esteves13, shporer14}.

\subsection{Reflected light and thermal emission}
\label{sec:ref}

As noted above, in this component of the orbital phase modulation we include two processes: light from the primary star reflected off the companion's atmosphere, and, thermal emission due to heating of the companion's atmosphere. Short-period systems are expected to have circular orbits and a low-mass companion is expected to be tidally locked where its rotation is synchronized with the orbit \citep{mazeh08}. Hence the orbital phase modulation of both reflected light and thermal emission are expected to be sinusoidal and show a maximum at superior conjunction and minimum at inferior conjunction (see \figr{scheme} green curve), while the amplitude depends on the difference in reflectivity (albedo) and surface temperature between the day-side hemisphere (facing the host star) and night-side hemisphere (facing away from the host star). We discuss deviations from this simplistic view at \secr{atm}.

Therefore we refer to the combined signal of these two processes as the atmospheric phase component, while the first two components described above (\secr{beam} and \secr{ellip}) are gravitational in origin. A simple description of this component is the following:
\begin{equation}
\begin{multlined}
\label{eq:atm}
\Aref \simeq \alref 0.1 \left( \frac{R_2}{a}\right)^{2}  \sin i 
= 57\ \alref \sin i \\
\times \left( \frac{M_1}{\msun}\right)^{-2/3} \left( \frac{P}{\rm day}\right)^{-4/3} \left( \frac{R_2}{\rjup}\right)^{2}    \rm ppm\ ,
\end{multlined}
\end{equation}
where the \alref\ coefficient is an order of unity coefficient that depends on the efficiency of atmospheric heat redistribution and reflectivity. 

For low-mass substellar companions to Sun-like hosts the companion is much cooler than the primary. Therefore this phase component is strongly dependent on wavelength, where reflected light dominates the optical phase curve and thermal emission becomes dominant at longer wavelengths, in the infrared (IR). In fact, IR phase curves are routinely being observed by the \is\ space telescope for short period gas giant planets \citep[e.g.,][]{knutson08, knutson12, lewis13, wong16}, where the amplitude is at the order of 0.1\%, about 1--2 orders of magnitude larger than the corresponding amplitude in the optical \citep[e.g.,][]{snellen09, mazeh10, esteves15, shporer15}.\\

All the equations above describing the various photometric amplitudes were derived while assuming the companion is of low mass and any nuclear burning it may have does not contribute to the light observed in the optical. In case the companion is luminous then it will generate its own phase curve modulations and the observed signal will be the superposition of the phase signals of both stars in the system. Since those signals are offset by half an orbit, the beaming and atmospheric components from the two stars will act to destruct each other while the ellipsoidal signal will be added constructively \citep{zucker07, shporer10}.

\begin{figure}[h]
\centering
\includegraphics[width=8.8cm]{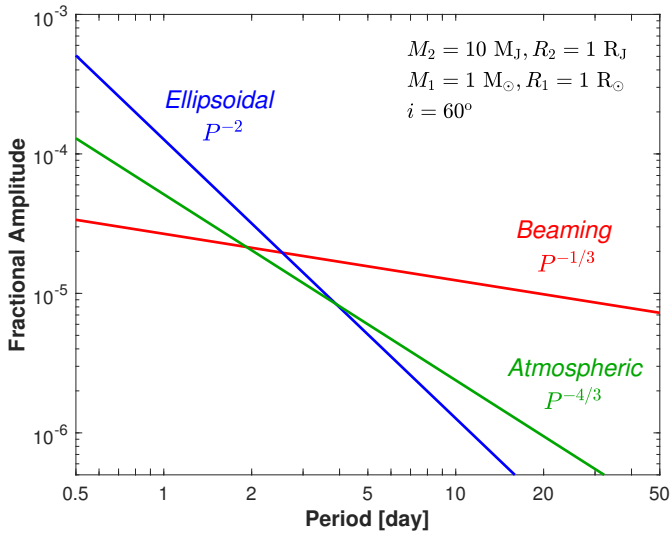}
\caption{Photometric amplitude, in relative flux, of the three phase components in optical phase curves as a function of orbital period, in log-log scale. The amplitudes are calculated using Equations~3, 7, and 9, for a hypothetical system where a 10 Jupiter mass and 1 Jupiter radius companion is orbiting a star with Solar properties in a circular orbit, viewed with an inclination angle of 60 degrees. The beaming, ellipsoidal, and atmospheric phase components are plotted and labeled in red, blue, and green, respectively, where the dependence on the orbital period, $P$, is also labeled. The figure shows that the beaming phase component decreases the slowest with increasing period.}
\label{fig:amp_per}
\end{figure}

\figr{amp_per} shows the amplitudes of the three phase components as a function of period for a hypothetical system with a 10 Jupiter mass and 1 Jupiter radius object orbiting a star identical to the Sun in a circular orbit. It is further assumed that the system has an orbital inclination angle of $i~=~60$~deg and it is observed in the optical. The figure shows that while all amplitudes decrease with increasing orbital period the beaming amplitude deceases the slowest as it has the weakest dependency on period.

\subsection{Other orbital phase processes}
\label{sec:other}

The processes described in the above three sections (Sec.~2.1--2.3) comprise the primary phase components expected to contribute to orbital phase curve photometric variability. We discuss here additional smaller contributions, induced by other processes or special circumstances.

\subsubsection{Multi planet systems}
\label{sec:multi}

When the primary star hosts more than a single companion, for example in a multi-planet system, then the phase curve signal becomes a superposition of the individual orbital signals. Since multi-planet systems tend to comprise of low-mass planets, at the Neptune mass level and below \citep{latham11} it is not likely that orbital signals of the individual planets will be identified. It might be possible, though, to measure the photometric signal when the signals from the individual planets superpose constructively. Such a cumulative photometric signal can be sensitive to the average atmospheric properties of the planets in the system, especially when their radii are known from transit measurements \citep{kane13, gelino14}. A similar approach was also applied to a multi planet system using RVs, aimed at studying the planets' mass and density \citep{weiss15}.

\subsubsection{Eccentric systems}
\label{sec:ecc}

When the orbit is non-circular the photometric signal becomes more complicated. For small eccentricity the phase curve becomes asymmetric and includes higher orbital harmonics. For high eccentricity most of the photometric signal becomes focused around the periastron (closest approach) phase. In these cases the light curve can look similar in shape to that of an electrocardiogram, earning these objects the nickname ``heartbeat stars" \citep{thompson12, shporer16b}. In many of these systems the stars show tidally excited stellar pulsations throughout the entire orbit, resulting from near resonances between multiples of the orbital frequency and that of asteroseismic modes \citep[e.g.,][]{kumar95, burkart12, fuller12}, therefore these systems are astrophysical laboratories for the study of stellar tides.

\subsubsection{Exo-Trojan asteroids and stacked phase curves}
\label{sec:ecc}

Trojan asteroids are asteroids co-orbiting with a planet at the orbit's Lagrangian points L4 and L5, located about 60 degrees ahead and behind the planet. In the Solar System Jupiter has several thousand known Trojans and some of the other Solar System planets have several known Trojans as well. Given their position along the orbit Exo-Trojans may show transits in the orbital phase curves of transiting planets, leading and trailing the transit by 0.17 in phase. However, an Exo-Trojan transit signal is expected to be small given their small size (with a non-spherical volume equivalent to that of a sphere with radius of up to 100 km, and typically much smaller). To try and overcome that challenge \cite{hippke15a} have stacked \ik\ phase curves of different systems. Such an approach allows to look for Trojans statistically, in a sample of systems, but not in individual systems. \cite{hippke15a} detect a small but statistically significant signal at the phase position where Trojans are expected for a sample of transiting planets on relatively long period beyond 60 days. This demonstrates the potential of the approach of phase curve stacking, which was also used in a similar way (although not exactly the same) by \cite{hippke15b} to look for exo-moons, and by \cite{sheets14} to look for the optical secondary eclipse of small planets.

\section{Astrophysical uses of optical phase curves}
\label{sec:astro}

\subsection{Probing the population of low-mass companions to hot stars}
\label{sec:hotstars}

\begin{figure}
\hspace{-8mm}
\includegraphics[scale=0.31]{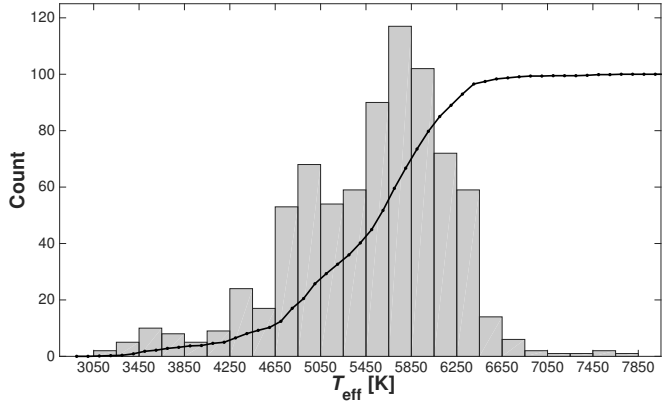}
\caption{\label{fig:teffhist}
Histogram of planet host stars effective temperature (\teff) for planets with dynamically measured mass, as reported on the NASA Exoplanert Archive \citep{akeson13}. The overplotted black line is the cumulative function, normalized to a maximum of 100 for visibility. The figure shows that while there is a gradual increase in the frequency of planet host stars with increasing \teff\ up to almost 6000~K, for hotter stars there is a sharp decrease with increasing \teff. That decrease is centered on the \teff\ range where the stellar convective region decreases dramatically, suggesting it is one of its causes. See text for further discussion. The histogram does not include V0391 Peg \citep{silvotti07}, with $\teff$~=~29300~$\pm$~500~K, for clarity, and KOI-55 \citep{charpinet11} where the detection of a planetary system was contested \citep{krzesinski15}.  
}
\end{figure}

The vast majority of the currently known exoplanet population is orbiting Sun-like stars and low-mass stars, with a spectral type range from mid-F to M. According to the NASA Exoplanet Archive \citep{akeson13} only 1.35~\% of planet host stars have $\teff >$ 6650~K, a fraction that deceases to only 0.48~\% for $\teff >$ 7050~K. \figr{teffhist} shows the distribution and the distribution cumulative function of known planet host stars \teff\ for planetary systems with a dynamically measured planet mass. Compared to a gradual increase in cool host stars frequency with increasing \teff\ the distribution shows a sharp decrease beyond roughly 6250~K. This matches the \teff\ region where there is a fast transition between stars with a convective layer and hot stars without it, suggesting it is one of the reasons for the sharp decrease.

Indeed there are differences between spectra of convective Sun-like stars and that of non-convective hot stars. For the latter the lack of a convective layer prevents angular momentum to be transported away via stellar winds, therefore hot stars retain their primordial angular momentum (spin) and remain fast rotators throughout their main sequence lifetime. The typical rotation rate, \vsini\ (where $V$ is the stellar equatorial rotation velocity and $I$ the stellar spin axis inclination with respect to the normal to the sky plane), for early F stars is a few 10 \kms, increasing to faster rotation rates for stars of earlier spectral types \citep[e.g.,][]{gray82, fekel97, wolff97, abt02}. The increased rotation rate of hot stars leads to rotationally broadened spectral lines, which in turn quickly degrades the precision of RV measurements, making it comparable to and larger than the RV amplitude the host star is expected to show due to an orbiting planet. Therefore, hot stars are inaccessible to high-precision RV measurements --- the most common method for dynamically measuring the mass of planetary companions --- leading to the poor knowledge of the planet population hot stars host, in sharp contrast to planets orbiting Sun-like hosts.

Another spectral difference making RV measurements more challenging for hotter stars, although with a more gradual dependence on \teff, is the decreasing number of spectral lines with increasing \teff, resulting from the increased number of ionized species in the stellar photosphere. 

There are of course other possible reasons that can contribute to the small number of known planets orbiting hot stars, like the increased stellar radius and mass leading to a decreased transit and RV signal, for the same planet. However, that should lead to a {\it gradual} decrease in the host stars \teff\ distribution and not the observed fast variation. Another possible reason can be a decrease in the number of hot stars that were monitored by exoplanet surveys. However, ground-based transit surveys are brightness-limited and monitor all stars in the field, and transit candidates are typically discarded only after spectroscopic follow-up data show it does not allow measuring the transiting planet's mass. Finally, the observed host star \teff\ distribution may reflect the intrinsic distribution, meaning planets orbiting hot stars are intrinsically less frequent than those orbiting Sun-like stars. However, testing this hypothesis requires efficient methods to detect that planet population and support statistical analyses.

Broad-band photometry is insensitive to the shape and number of spectral lines. Therefore, since orbital phase curves are sensitive to the planet's mass they can potentially be used to probe the planet population orbiting hot fast-rotating stars. In addition, as a population hot stars are photometrically quiet, making them more accessible to studying a low-amplitude photometric orbital signal.

A good example of the above is the Kepler-13A (KOI-13, KIC~9941662) system, containing an A type host star and a short period (1.76~day) massive gas giant planet \citep{shporer11, szabo11, mazeh12, mislis12, esteves13, placek14}. The transit was detected in \ik\ data but the star's fast rotation ($\vsini = 76.6 \pm 0.2 $~\kms, \citealt{johnson14}) prevents obtaining RVs precise enough to detect the star's orbital motion induced by the transiting companion, as shown by \cite{santerne12} who were only able to place an upper limit on the RV variability. Still, the latter proved that the companion's mass is planetary. In fact, Kepler-13 is a four body system. It is a visual binary system composed of two almost identical A stars, Kepler-13A and Kepler-13B, with an angular separation of 1.15\arcsec \citep{adams12, law14, shporer14}. Kepler-13A is the transiting planet host and Kepler-13B hosts a non-eclipsing stellar companion (Kepler-13BC) on a 65.8~day orbit \citep{santerne12, johnson14}.

\begin{figure}
\hspace{-5mm}
\includegraphics[scale=0.5]{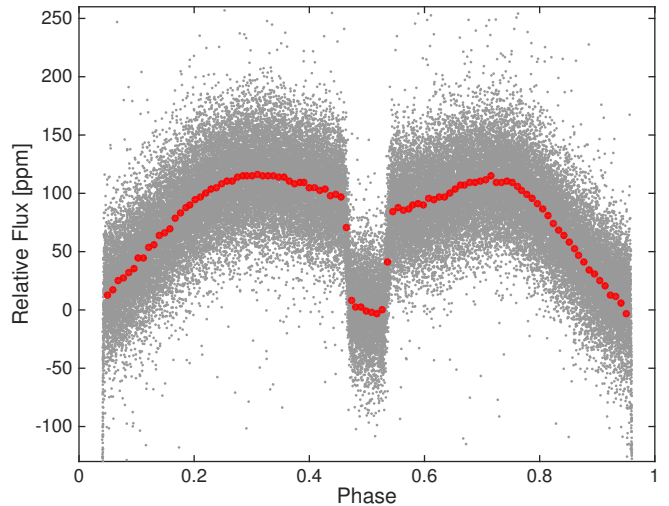}
\caption{\label{fig:kep13}
\ik\ phase folded light curve of Kepler-13. The $y$-axis shows the relative flux in parts per million (ppm) relative to the mean flux within the secondary eclipse, seen here at phase~0.5, when the planet is fully eclipsed by the star. The figure is centered on the phase modulation so the transit, at phase zero, is out of scale. \ik\ long cadence data is in gray points, overplotted by a binned light curve in red circles (bins errors are comparable to marker size and not plotted). The orbital phase modulation is clearly seen in the un-binned data, with a full variability amplitude at the order of 100 ppm. The data plotted in the figure is not corrected for the blending of the planet host star with its visual binary companion \citep{shporer14}. 
}
\end{figure}

\figr{kep13} shows the phase folded \ik\ light curve of Kepler-13. The figure focuses on the phase curve variability and secondary eclipse so the transit is out of the figure's $y$-axis scale. The light curve is plotted in relative flux, relative to the flux during the secondary eclipse.

The Kepler-13 phase curve was analyzed by several authors \citep{shporer11, mazeh12, mislis12, esteves13, placek14, shporer14, esteves15} who used the amplitudes of the two gravitational phase components, the beaming effect and the tidal ellipsoidal distortion, to estimate the transiting companion's mass. There are several factors that contribute to the uncertainty on the derived mass using this method:

\begin{itemize}

\item {\bf Photometric data quality:} This determines how precise the phase curve shape is measured and in turn the precision of the amplitudes of the beaming effect and tidal ellipsoidal distortion. The data quality is typically determined by the photometric precision of individual measurements, the total amount of measurements, and their time span. The exposure time does not directly affect the overall data quality, since phase curves show a relatively slow photometric variation, at the time scale of the orbit, which is a few orders of magnitude longer than the typical exposure times used in photometric surveys (e.g., 1 and 30 minutes for \ik\ and K2).

\item {\bf Host star physical understanding:} The directly measured quantities, the photometric amplitudes of the phase curve components, depend on the planet's mass and also the host's stellar parameters, specifically stellar mass, radius, and temperature (see Equations 3 -- 8). Therefore, measuring the planet's mass requires a physical understanding of the host star. This is also the case in other methods designed to measure the planet's properties, like the measurement of the planet mass with the RV method and the planet radius with the transit method.

\item {\bf Blending:} When a target's PSF is fully or partially blended with that of another star the target's photometric variability is diluted, decreasing the measured variability amplitude (and may even lead to confusion as to the true source of the variability). This in turn leads to a downward bias in the measured planet's mass. Photometric surveys are typically designed to allow some dilution in order to obtain a wide field of view and monitor a large number pf stars (e.g., \ik\ pixels stretch 3.98\arcsec\ and the PSF is a few pixels wide, while the FOV area is $105\deg^2$; \citealt{koch10}). Therefore when measuring photometric variability one always should consider the possibility of blending. For transit surveys this is routinely done by looking for variation in the center of light position \citep{bryson13} and carrying out high angular resolution imaging of the targets \citep[e.g.,][]{everett15, baranec16, furlan17}. The uncertainty on the amount of blending then propagates into the measurement of the corrected, or de-blended, photometric variability, and from there to the planet's mass estimate.

\item {\bf Incomplete orbital phase curve understanding:} The description given in \secr{processes} of the various phase components contains some approximations and assumptions. Those may not always hold, leading to a possible bias in the photometric amplitudes measurement. In addition, as described in \secr{other} there may be additional processes or higher order effects impacting the phase curve shape. It is difficult to estimate how this contributes to the planet mass uncertainty, especially since this is a young method. Some examples of the incomplete understanding of phase curves and deviations from the simplistic picture painted in \secr{processes} are discussed in \secr{massdiscrep}.

\end{itemize}


Specifically for Kepler-13Ab, the target is relatively bright (\kp~=~9.96~mag) and photometrically quiet, so the minute orbital modulations are clearly seen and precisely measured (see \figr{kep13}). Still, the planet mass has a wide range of estimated values in the literature, from about 5 to 12~\mjup\ \citep{shporer11, mazeh12, mislis12, esteves13, placek14, shporer14, esteves15}. This is due to the latter three factors described above. 

The stellar parameters have changed significantly from the initial estimate in the Kepler Input Catalog \citep[KIC;][]{tbrown11}, based on wide band photometry, to the estimate based on the target's spectrum \citep{shporer14}. That is not surprising since the KIC values were designed to be accurate for Sun-like stars, and have decreased accuracy for hot stars and low-mass stars \citep{tbrown11}.

The amount of blending, and the blending uncertainty, have also evolved considerably when the spectroscopic and imaging data became available. \cite{shporer14} estimated the blending and the blending uncertainty based on the spectra of the two visual binary components, Kepler-13A and Kepler-13B, while integrating over the \ik\ bandpass.

Finally, a few authors have noticed that the two planet mass estimates, one based on the beaming effect amplitude and the second based on the tidal ellipsoidal distortion amplitude, are statistically discrepant \citep{shporer11, mazeh12, shporer14}. This points to an incomplete understanding of the orbital phase curve and is discussed in detail in \secr{massdiscrep}. 

\cite{shporer14} have attempted to account for all contributions to the planet mass uncertainty as described above, and derived an estimated range of 4.94~--~8.09~\mjup, covering the two discrepant mass estimates based on the beaming effect and tidal ellipsoidal distortion amplitudes.

The discovery and mass measurement of planets orbiting hot stars is important not only on its own right, but also because it allows the opportunity for more detailed follow-up studies of planets in a different environment. At the time of discovery Kepler-13A was the hottest and fastest rotating star known to host a short period gas giant planet and Kepler-13Ab was the gas giant planet with the hottest known atmosphere. This has motivated the investigation of the planet's atmosphere using both ground-based and space-based observatories \citep{shporer14, beatty16}, and the measurement of the host star obliquity \citep{barnes11, johnson14}.

To complete the discussion we list below other photometric methods capable of probing low-mass companions to hot stars and measuring the companion mass, thereby going beyond the reach of the RV method. We note that all methods below are sensitive to companions at wide orbits, from about 1~au and beyond, while phase curves are sensitive to companions on short orbital periods, at a few days and weeks. 

\begin{itemize}

\item {\bf Stellar pulsation phase modulation.} This method uses the time delay of photometrically-measured stellar pulsations due to light travel time along the orbit \citep{silvotti07, shibahashi12, balona14, murphy14, murphy15, kutz15, murphy16a, murphy16b}. It requires stars with stable pulsations, like delta Scuti ($\delta$~Sct) pulsators which are A stars positioned on the H-R diagram where the instability strip intersects with the main sequence. The measured signal increases with increasing orbital period as that increases the number of stellar pulsations per unit orbital phase.

\item {\bf Astrometry.} Precise astrometric measurements are sensitive to the sky-projected orbital motion of the host star as it orbits the system's center of mass. As this method measures a two-dimensional projection of the orbital motion, as opposed to a one-dimension projection with the RV method, it gathers more information about the orbit. This translates to a measurement of both the companion mass $M_2$ and orbital inclination $i$, instead of the so-called minimum mass, $M_2 \sin(i)$, measured with the RV method. \citet[][see also \citealt{sozzetti01, sozzetti14, ranalli17}]{perryman14} have done a thorough analysis of the predicted yield of the Gaia mission data \citep{perryman01, gaia16} and found it is expected to be sensitive to low-mass companions down to the Jupiter mass level at orbits with semi-major axes of a few au.

\item {\bf Gravitational microlensing.} The directly measured quantity is the system's mass ratio, measured from an anomalous transient photometric signal generated when the light from a distant star (the source) passes within the gravitational field of a star-planet system (the lens). Although it provides less information about the system than methods based on detecting an orbital periodic signal, it is sensitive to low-mass planets, down to about Earth mass using ground-based surveys \citep[e.g.,][]{gaudi12} and expected to reach below 0.1~Earth mass in space-based surveys \citep{spergel15}. Similarly to the other methods mentioned above, microlensing is sensitive to planets on wide orbits, at about 1~--~10 au.
One caveat of using photometric microlensing surveys to detect low-mass companions to hot stars is that they are volume limited surveys, hence the fraction of hot stars is small following the initial mass function and the decreasing stellar lifetime with increasing stellar mass. For example, the NASA Exoplanet Archive currently lists 26 planet host stars detected by microlensing and with a mass uncertainty below 0.15~\msun. The mass of all those host stars is measured to be below 1.2~\msun, and for $\approx90\%$ of them the mass is below 0.75~\msun.


\end{itemize}

\subsection{Detecting non-transiting systems}
\label{sec:noneclipse}

Photometric phase modulations are of course present also when the system does not show transits (or eclipses). Therefore, they can be used to detect companions (almost) independently of the system's orbital inclination, similarly to RV surveys. The geometric probability of an eclipse, or transit, in a binary or star-planet system equals the sum of the two objects' radii divided by the distance between them at conjunction. In a circular orbit the distances during superior and inferior conjunctions are the same, while they differ in an eccentric orbit leading to different probabilities between primary eclipse and secondary eclipse. For a Sun-like host and a low-mass companion in a circular orbit this probability is about 10\% for a 4-day orbit, dropping to 5\% at 10 days and below 3\% at 30 days. Therefore the vast majority of systems are not eclipsing, motivating designing methods targeting non-eclipsing systems. 

On the other hand, photometric surveys simultaneously monitor many stars. The \ik\ mission for example has simultaneously monitored close to 200,000 stars \citep[e.g.,][]{borucki16, coughlin16, twicken16}. The combination of sensitivity to systems with a wide range of orbital inclination and the simultaneous monitoring of a large number of stars is a fundamental advantage this innovative approach has over the traditional RV and transit surveys. 
In addition, using photometry allows access to host stars not accessible to RV surveys, like hot and fast rotating stars (as discussed in the previous section).

This approach was described in detail by \cite{faigler11} who presented the BEER algorithm, designed to detect light curves with photometric modulations that are likely to be orbital phase modulations of non-eclipsing systems. The algorithm is based on the fact that each of the three phase components (See \secr{processes}) can be approximated by a sinusoidal modulation. As shown in \figr{scheme}, assuming phase zero to be at inferior conjunction the beaming and atmospheric components are a sine and cosine modulations, respectively, at the orbital period, and the ellipsoidal component is a cosine at the first harmonic of the orbital period. The two cosine components (atmospheric and ellipsoidal) are expected to have negative coefficients (or amplitude) and the sine component (beaming) a positive coefficient.

The algorithm consists of a period search, the orbital period, where for each trial period the phase folded light curve is fitted with a double harmonic model, composed of a sinusoidal component at the trial period and a second sinusoidal component at the first harmonic \citep[e.g.,][]{shporer06, shporer07}. 

The fitting is done in two steps. In the first step a linear 5-parameter model is fitted: 
\begin{equation}
\begin{multlined}
\label{eq:model}
f(t) = a_0 + a_{1c}\cos(\frac{2\pi t}{P}) + a_{1s}\sin(\frac{2\pi t}{P}) \\
 + a_{2c}\cos(\frac{2\pi t}{P/2}) + a_{2s}\sin(\frac{2\pi t}{P/2}),
 \end{multlined}
\end{equation}
where $t$ is time with an arbitrary zero point. Since the phase curve is not expected to have a sine component at the first harmonic but only a cosine component, after the fit is done the time $t$ is set to $t'$, a time with a zero point that brings $a_{2s}$ to zero while $a_{2c}$ is negative (see more details in \citealt{faigler11}). 

Then in the second step of the fitting a linear 4-parameter model is fitted:
\begin{equation}
\begin{multlined}
\label{eq:model}
f(t') = a'_0 + a'_{1c}\cos(\frac{2\pi t'}{P}) + a'_{1s}\sin(\frac{2\pi t'}{P}) \\
 + a'_{2c}\cos(\frac{2\pi t'}{P/2}).
\end{multlined}
\end{equation}
In this fit the coefficients $a'_{1c}$, $a'_{1s}$, and $a'_{2c}$ correspond to the atmospheric, beaming, and ellipsoidal phase components, respectively. 

The resulting coefficients are then tested. First, $a'_{1c}$ is expected to be negative and $a'_{1s}$ is expected to be positive. Second, as shown by Equations 4 and 7 both the beaming and ellipsoidal photometric amplitudes depend linearly on the companion mass. Therefore, assuming $M_2 \ll M_1$, their ratio is a quantity that does not depend on the companion's mass, or any other intrinsic property of the companion, but only on the host star properties and the orbital parameters:
\begin{equation}
\begin{multlined}
\label{eq:ratio}
\mathcal{R} \equiv \frac{\Aell}{\Abea} = \\ 5\frac{\alell}{\albea}  \left( \frac{R_1}{\rsun}\right)^{3} \left( \frac{M_1}{\msun}\right)^{-4/3} \left( \frac{P}{\rm day}\right)^{-5/3} \sin i .
\end{multlined}
\end{equation}
Therefore, the observed ratio ($a_{2c} / a_{1s}$) is compared to the expected ratio $\mathcal{R}$, resulting in a likelihood $\mathcal{L}(P)$ of the fitted beaming and ellipsoidal modulations to be the result of an orbiting companion at period $P$. The value assigned to each period $P$ in the BEER periodogram is the model's goodness-of-fit multiplied by the  likelihood $\mathcal{L}(P)$, where the goodness-of-fit is taken to be the ratio of the model rms and the residuals rms \citep{faigler11}.

\begin{figure*}
\begin{center}
\includegraphics[scale=0.51]{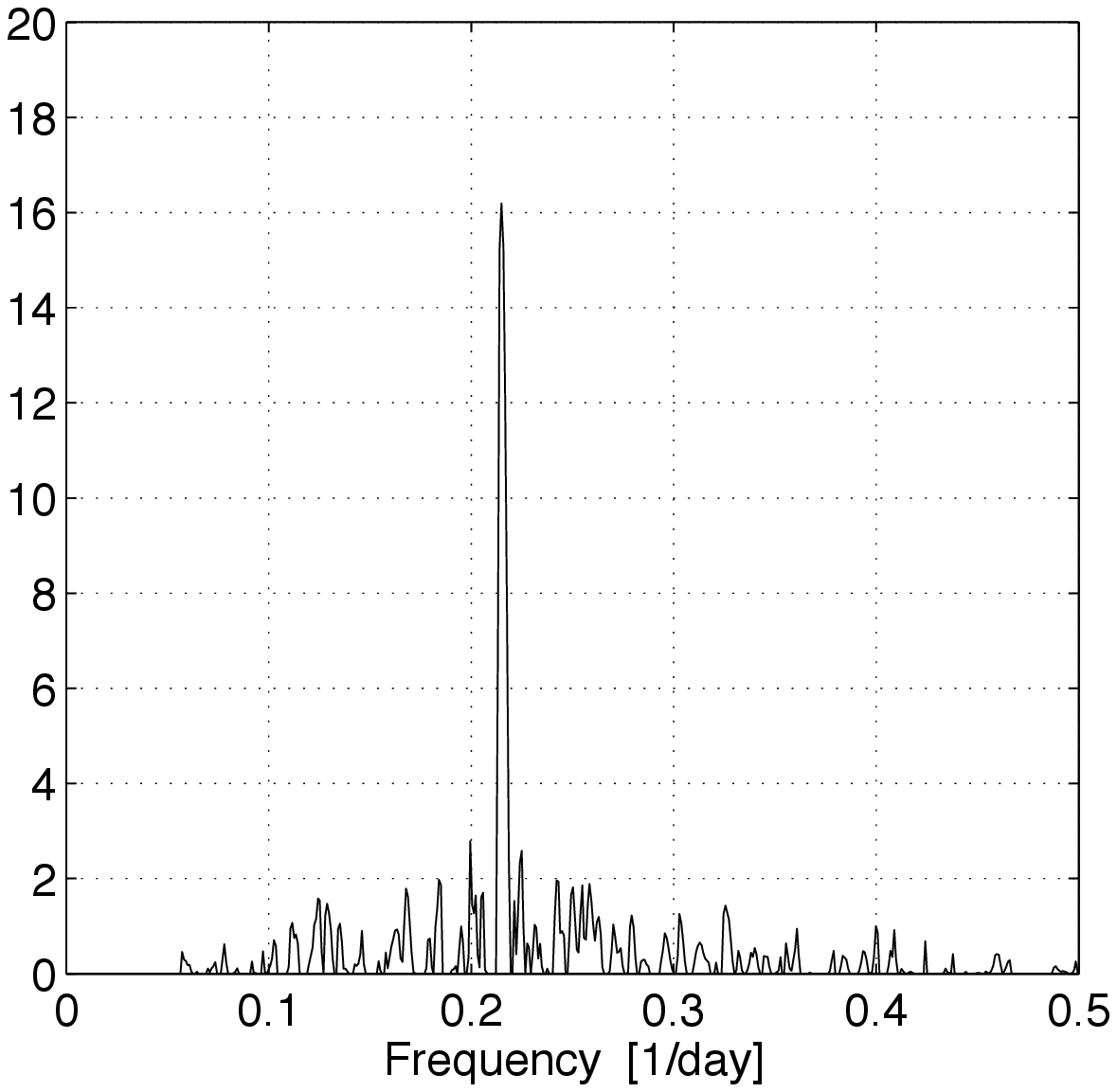}
\includegraphics[scale=0.51]{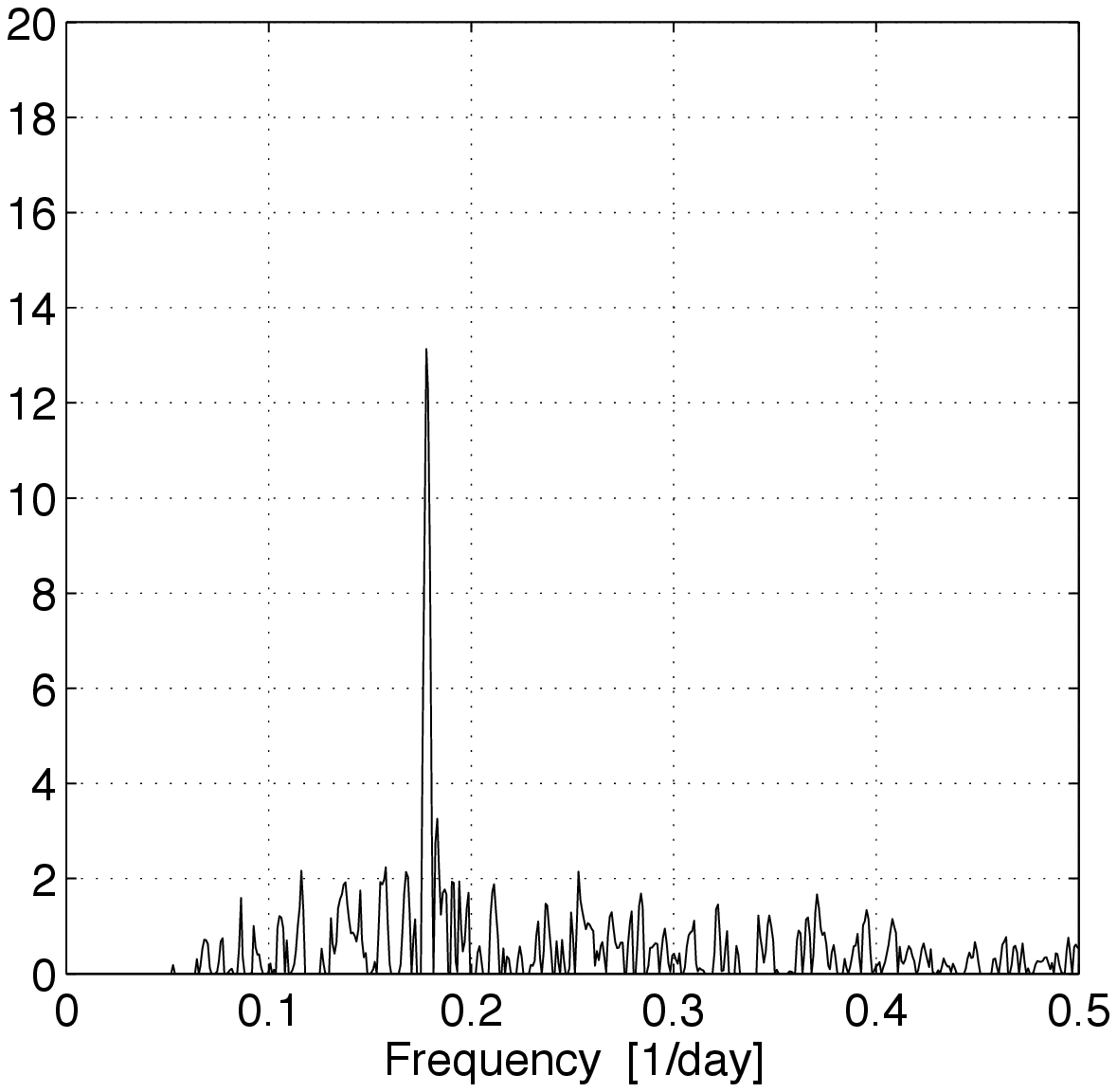}\\
\includegraphics[scale=0.51]{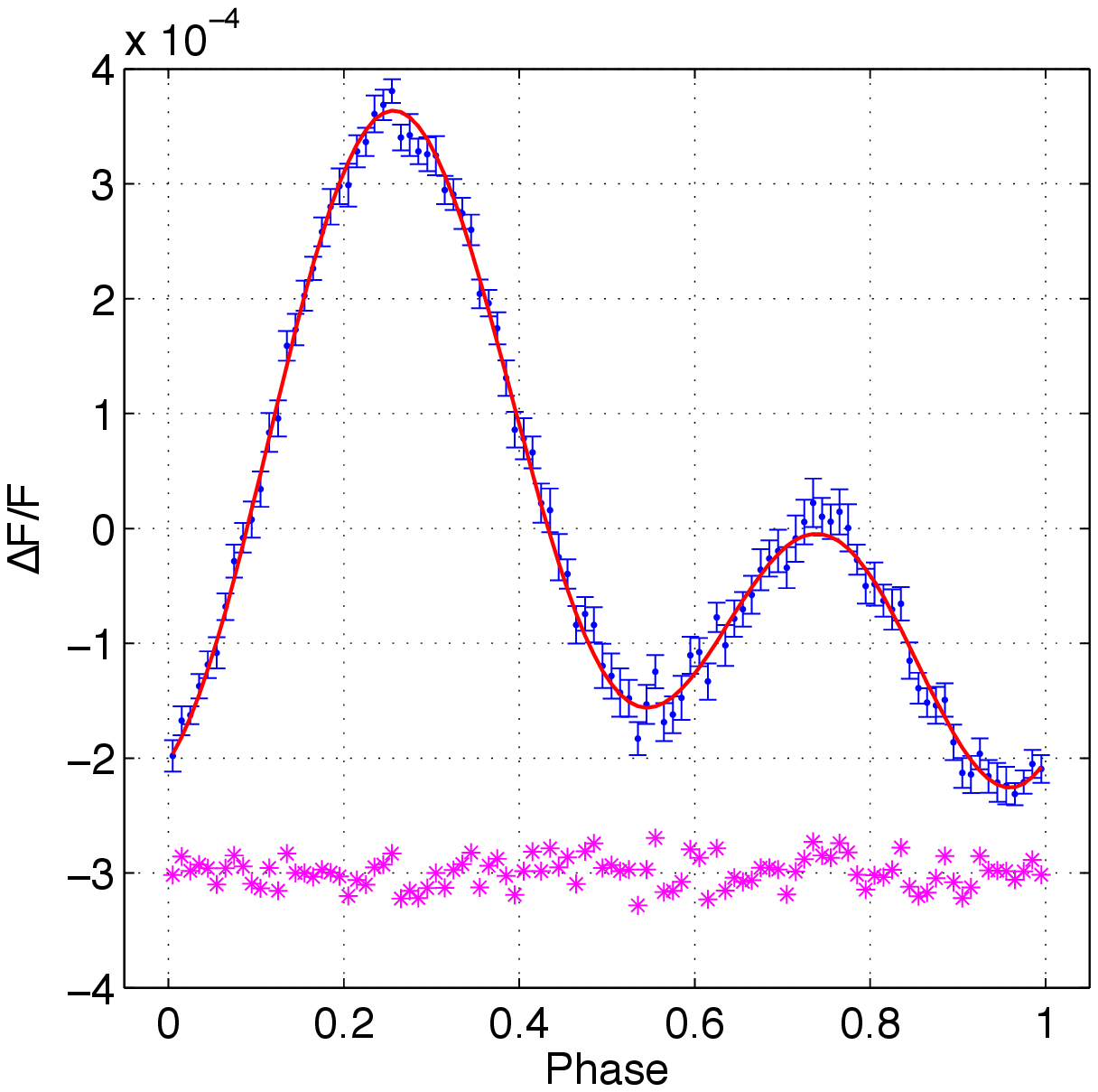}
\includegraphics[scale=0.51]{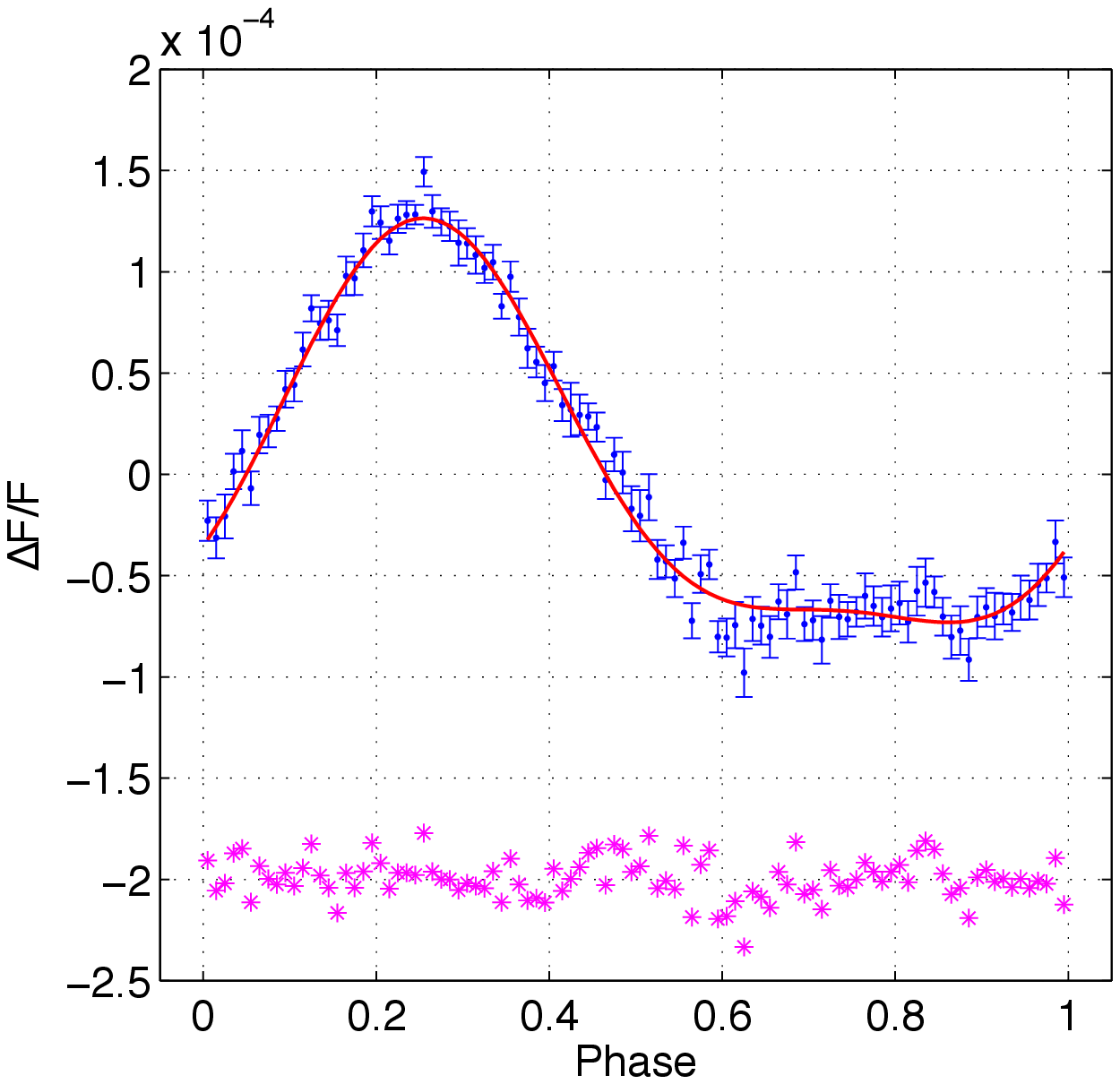}\\
\hspace{-4mm}
\includegraphics[scale=0.81]{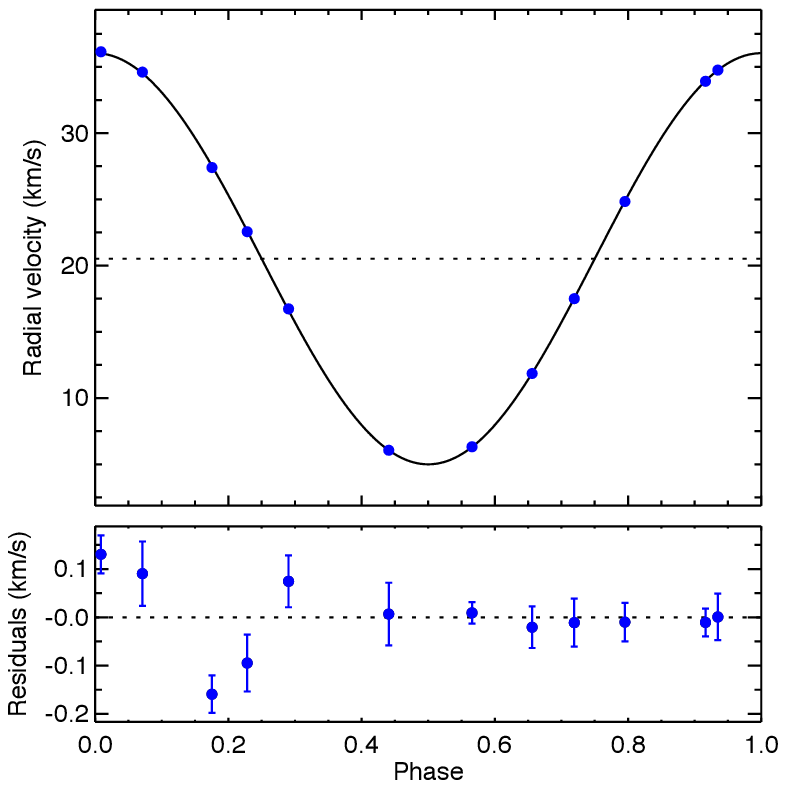}
\hspace{6mm}
\includegraphics[scale=0.81]{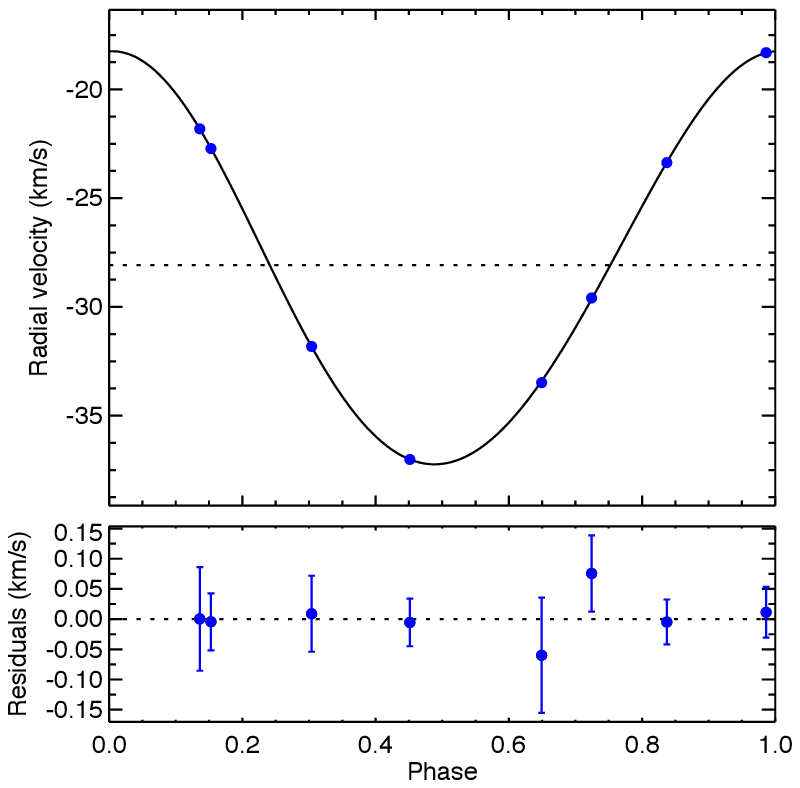}
\caption{\label{fig:beer}
Two examples of a detection of a non-eclipsing companion by applying the BEER algorithm \citep{faigler11} to \ik\ data. KIC 9512641 ($P$ = 4.65 day) is shown in the left panels and KIC 8016222 ($P$ = 3.49 day) in the right panels. Figures adopted from \cite{faigler12}. {\it Top:} BEER periodograms. {\it Mid:} Phase folded light curves (blue) in relative flux, overplotted by the BEER model (red line). Residuals are shown at the bottom of each panel. {\it Bottom:} Phase folded radial velocity curves (measurements in blue, fitted Keplerian model in black, residuals at the bottom). The top two rows (periodograms and phase folded light curves) present the photometric detection of a non-eclipsing companion, and the bottom row shows the RV confirmation. The minimum mass of KIC 9512641 secondary is $147\pm10$~\mjup\ and of KIC 8016222 it is $90\pm6$~\mjup\ \citep{faigler12}.
}
\end{center}
\end{figure*}

\figr{beer} shows two examples of using the BEER algorithm to detect a non-eclipsing companion in \ik\ data, for KIC 9512641 (left panels; $P$ = 4.65 day) and KIC 8016222 (right panels; $P$ = 3.49 day), taken from \cite{faigler12}. The figure shows the BEER periodograms (top row panels), \ik\ phase folded light curves (mid row panels), and RV confirmation (bottom row panels). A visual inspection of the \ik\ phase folded light curve of KIC 9512641 (mid left panel) shows it has a double peak sinusoidal structure, hence it is dominated by the ellipsoidal modulation. The two maxima are not equal as a result of the beaming modulation, and the two minima are not equal due to the atmospheric phase component. The light curve of KIC 8016222 (mid right panel) shows a somewhat different shape, where the second peak (at phase 0.75) is completely suppressed since the beaming component is comparable in amplitude to the ellipsoidal component. These two examples show that phase curves induced by an orbiting companion do not all have a similar schematic shape, like for example eclipse or transit light curves.


Close examination of \eqr{ratio} shows that it can be used to estimate the orbital inclination, since all other parameters in that equation are either directly measured ($P$, \Abea, \Aell) or derived from characterizing the host star ($M_1$, $R_1$, \albea, \alell). This results from \Abea\ being dependent on $\mtsini$ (see \eqr{abeam3}) while \Aell\ depends on $M_2 \sin^2i$ (see \eqr{ellip}), allowing in principle to measure $M_2$ and $\sin i$ independently in a non-eclipsing system. However, in practice this turns out to be difficult since it requires precise estimates of the host star's radius and mass as well as the \albea\ and \alell\ coefficients. It is also hindered by deviations of the phase components from the simplistic model presented in \secr{processes}, as discussed in \secr{atm} and \secr{massdiscrep}.

\begin{figure*}
\begin{center}
\includegraphics[scale=0.75]{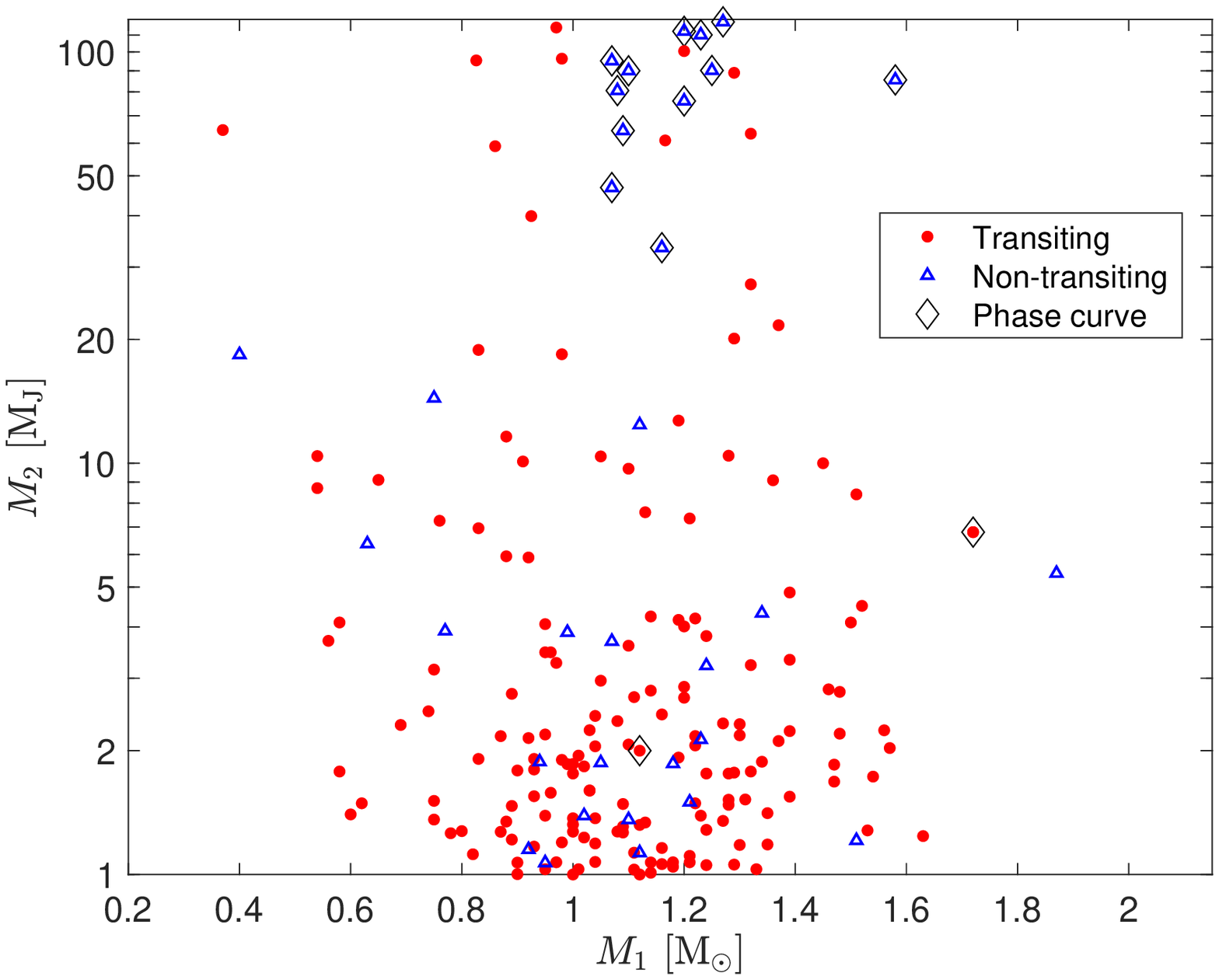}
\caption{\label{fig:m2m1}
Companion mass (in Jupiter mass; logarithmic scale) as a function host star mass (in Solar mass; linear scale), for short period systems ($P~<~30$~d) with companions in the 1~--~120~\mjup\ range, including massive planets, brown dwarfs, and low-mass stars. Transiting companions are marked by red circles and non-transiting companions where only a lower limit on their mass is known (minimum mass, \mtsini) are marked by blue triangles. Objects marked by a diamond were detected through their photometric phase modulations by the BEER algorithm, or have phase modulations that allow detecting them even if they were not transiting. Those include Kepler-76b (bottom part of the diagram; \citealt{faigler13}), Kepler-13Ab (right part of the diagram; \citealt{shporer14}), and 12 objects, detected using BEER, at the top part of the diagram (2 from \citealt {faigler12} and 10 from \citealt{talor15}) of which some are brown dwarf candidates and others are close to the hydrogen burning mass threshold (companion minimum mass was not reported by \citealt{talor15} and we derived it here from their reported orbital period, primary RV amplitude, and primary mass). Error bars were not plotted to avoid cluttering the figure. The typical precision on $M_1$ and $M_2$ is 5--15\%. The figure shows the decreased occurrence of companions in the brown dwarf mass range compared to planets (and stellar companions, although those are not shown), and that using phase curves to detect short period companions is already starting to fill that void. The figure also shows the decreased occurrence of companions around low-mass stars, believed to be astrophysical since the RV orbital amplitude increases with increased companion mass and decreased host star mass, and the decreased occurrence of companions around massive stars, which is at least partially an observational bias since massive stars are not accessible to high precision RV measurements at the level required to measure the companion's mass.
Data taken from the NASA Exoplanet Archive and the literature. 
}
\end{center}
\end{figure*}

The expected sensitivity of the \ik\ data to orbiting companions using the BEER approach was estimated by \cite{faigler11} and \cite{shporer11} in different ways. They both reach a similar conclusion, when extrapolating to the entire \ik\ data, that this method will be sensitive to short period companions, within $P \approx 10$ days, with mass down to a few Jupiter mass. This encompasses a region in $P$ and $M_2$ parameter space that is known to be intrinsically sparsely populated, including short period massive planets \citep[e.g.,][]{bakos07, bakos12} and the brown dwarf desert \citep[e.g.,][]{grether06, deleuil08, bouchy11, bayliss17}. This is visually seen in \figr{m2m1}, showing the companion mass $M_2$ against the primary mass $M_1$ for systems with $P~<~30$~day and a companion mass in the range of $1~<~M_2~<~120$~\mjup\ including massive planets, brown dwarfs, and small stars. It is visually clear that the occurrence rate drops significantly above a few Jupiter mass, and even more so above $\approx 10$~\mjup.

Therefore, the BEER approach is sensitive to an intrinsically rare population of objects, of which RV and transit surveys over the last few decades have yielded only about two dozen (\figr{m2m1}). That, combined with the advantage of this approach over RV and transit surveys, as discussed earlier, makes detecting brown dwarfs and massive planets at short periods one of the primary scientific motivations for applying the BEER approach to the large number of high quality light curves obtained by space-based photometric transit surveys. Increasing the small known sample of those objects is important for the study of formation and orbital evolution processes of both stars and planets \citep[e.g.,][]{armitage02, bate09, jumper13, ma14}.

Another scientific motivation is the detection of another rare population, that of short period compact object companions to regular stars \citep[e.g.,][]{cowley92, mcclintock06, shporer07, orosz07}. A compact object companion is expected to induce relatively large beaming and ellipsoidal photometric modulation amplitudes while not showing an atmospheric phase component, and, a compact object will not appear in the optical spectrum of the system.

Beyond detecting rare objects, the BEER approach can detect large samples of short period stellar binaries from a single survey (e.g., CoRoT and \ik), leading to well defined samples. Those can provide input to statistical studies of that population, for example its period and mass ratio distributions \citep[e.g.,][]{duquennoy91, mazeh92, raghavan10, talor15}. 

BEER detections are considered as candidates and require RV follow-up, to check that the photometric signal is due to an orbiting companion, measure the orbit and derive the companion's mass (or mass ratio) independently of the photometry. Similarly to RV follow-up of transit candidates, the orbital ephemeris is known from photometry which significantly decreases the number of RVs needed to measure the orbit.

The RV confirmation is essential since the photometric detections are prone to astrophysical false positives, where the detected photometric modulation is not due to an orbiting companion. Such false positive scenarios include:

\begin{itemize}

\item {\bf Stellar activity:} The combination of stellar rotation and non-uniform spots surface distribution results in sinusoidal photometric modulations at the rotation period \citep[e.g.,][]{hartman11, irwin11, mcquillan14}. The shape of activity-induced modulations can mimic the shape of a phase curve modulation induced by an orbiting companion as the two phenomena are sinusoidal and overlap in period and photometric amplitude. However, as the spots evolve the photometric modulation amplitude and phase vary, unlike the constant orbital signal. Therefore, identifying this false positive scenario requires looking for changes in the periodic signal over time. This can be done for example by comparing results from different subsets of the data or using a wavelet-based approach \citep{bravo14}.
Since the level of activity depends on the star's convective layer size this false positive scenario is expected to occur for M to late-F type stars and not likely to occur for early-F type and hotter stars. 

\item {\bf Seismic pulsations of evolved red giant stars:} While asteroseismic pulsations of main sequence stars are at a timescale of minutes \citep[e.g.,][]{chaplin11, chaplin13, chaplin14}, seismic pulsations of evolved red giant stars are at a time scale of days, following their larger radii and lower density, and they show sinusoidal modulations with amplitudes similar to that of orbital modulations \citep[e.g.,][]{hekker09, davies16, sharma16}. Therefore, such pulsations can be a source of a false positive scenario. Possible ways to identify this scenario, or at least flag BEER detections as suspicious, are to look for a multi periodic signal (following radial and non-radial pulsation modes), and examine the stellar properties.

\item {\bf Non-seismic stellar pulsations:} Some types of pulsating stars show sinusoidal periodic light curve variability with a similar period and amplitude to that of orbital modulations. These include for example gamma Doradus ($\gamma$ Dor) pulsators, with a period of about 1 day driven by convective blocking \citep[e.g.,][]{guzik00, balona11a}, and $\delta$ Sct pulsators with a period of several hours driven by variations in He II ionization \citep[e.g.,][]{balona11b, balona15}. Similarly to the above, possible ways to identify this scenario are looking for a multi periodic signal, following radial and non-radial pulsation modes, and examine the stellar properties, as each class of pulsating stars has specific properties. In the examples above both types are located in the HR diagram close to the intersection of the instability strip with the main sequence \citep[e.g.,][]{dupret04, Grigahcene10, kahraman16}. $\delta$ Sct stars are typically slightly hotter than $\gamma$ Dor stars although some stars show both pulsation patterns \citep[e.g.,][]{Grigahcene10, balona11b, balona15, bradley15}.

\end{itemize}

In addition to the above, as mentioned in \secr{hotstars} blending decreases the observed photometric variability amplitude which can bias the interpretation of the detected signal. 

A systematic search through available light curves and RV follow-up of BEER candidates is ongoing. \cite{faigler12} have detected 7 non-eclipsing low-mass companions, two of which close to the hydrogen burning mass threshold (see \figr{m2m1}). \cite{talor15} have identified 70 binary companions using the AAOmega multi-fiber spectrograph \citep{lewis02}, and another program using the WIYN/Hydra multi-fiber spectrograph is ongoing \citep{shporer16a}. The study of \cite{talor15} is the first demonstration of using the BEER approach and RV follow-up for detecting a large sample of short period binaries for statistical studies, and, it detected two brown dwarf candidates along with a few other objects close to the minimum mass needed for hydrogen burning (see \figr{m2m1}). 

An especially interesting discovery was done by \citet{faigler13} who discovered Kepler-76b, a $2.00~\pm~0.26\ \mjup$ transiting planet detected through the systems's phase curve modulation when applying BEER to \ik\ light curves. This discovery demonstrates the potential of the BEER approach for detecting massive planets.

\figr{m2m1} shows how the first RV follow-up campaigns mentioned above \citep{faigler12, talor15} are already starting to increase the amount of known objects belonging to the intrinsically rare population of short period low-mass companions. These first RV follow-up campaigns are aimed also at obtaining better understanding of the false positive scenarios of this young approach in order to improve its success rate at identifying low-mass companions. \cite{talor15} showed that for high quality candidates (referred by \citealt{talor15} as Priority 1 candidates) the false positive rate is $< 50\%$, and improves further when considering candidates beyond a certain period and companion minimum mass threshold. That rate is expected to improve in the future as better understanding of this method and the false positive scenarios is gained.

\subsection{Atmospheric characterization}
\label{sec:atm}

The atmospheric phase curve component provides a one dimensional longitudinal map of the companion surface brightness (assuming tidal locking). Therefore it is a rich source of information about the companion's atmosphere. In comparison, a secondary eclipse (when the companion moves behind the primary and is completely occulted) measures only the brightness of the companion's day-side hemisphere, although it is complementary to the information retrieved from a phase curve.

In the IR, orbital phase curves of short period gas giant planets (so called hot Jupiters) have been obtained by \is\ over the last decade, and their morphology is used for constraining the planetary atmosphere dynamics and composition \citep[e.g.,][]{knutson07, knutson12, zellem14, wong16}. 

One of the prominent features identified in IR phase curves is a phase shift between the phase of maximum flux and superior conjunction (phase of secondary eclipse for eclipsing systems), where the maximum flux appears before superior conjunction \citep[e.g.,][]{knutson07, knutson09, crossfield10, zellem14}, indicating that the atmosphere's hottest longitude is located eastward of the substellar point. This phase shift depends on the interplay between the time scales of advective and radiative processes. Those determine the fraction of the heat irradiated on the day-side hemisphere that is reradiated away, the fraction that gets deposited into the planet, and the efficiency of heat distribution around the planet from the day-side to the night-side hemispheres. 

The observed phase shifts in the IR are explained by theoretical models put forth before the IR observations were made \citep[e.g.,][]{showman02, cho03, fortney06}, where fast winds along the planet's equator transport heat eastward from the substellar point, generating a so called hot spot. That hot spot rotates into the observer's view \emph{before} the substellar point therefore causing the phase of maximum flux (in the IR) be earlier than the superior conjunction.

While in the IR the atmospheric processes completely dominate the phase curve, with an amplitude at the order of 0.1\% in relative flux for hot Jupiter systems, in the optical the atmospheric phase component amplitude is typically at a similar order of magnitude as the other two gravitational phase components. Hence all phase components need to be accounted for, either by fitting for their amplitudes or fixing them based on external information such as the planet mass (or the system mass ratio) derived from an RV-measured Keplerian orbit. Such analysis was done by several authors for a sample of known transiting hot Jupiters \citep{esteves13, esteves15, angerhausen15, vonparis16}, and a phase shift of the atmospheric component was identified in some of the phase curves.

\begin{figure*}
\begin{center}
\includegraphics[scale=0.42]{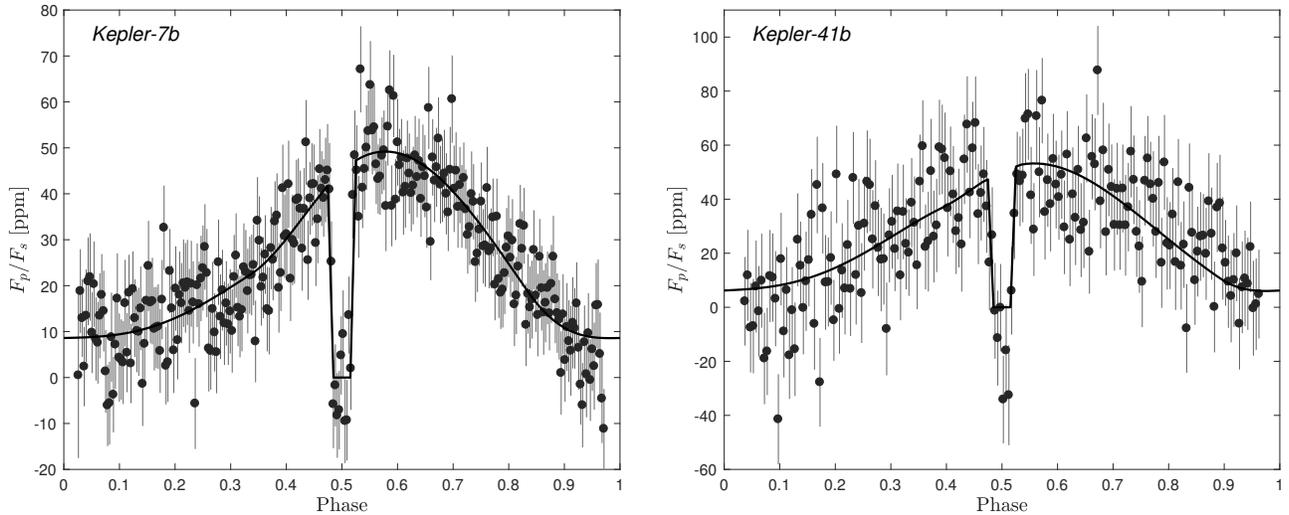}
\caption{
\label{fig:phaseshift}
Two examples of optical planetary phase curves showing maximum flux at a \emph{later} phase than superior conjunction (secondary eclipse, phase 0.5). The panels show the phase folded and binned \ik\ light curves of Kepler-7b (left; figure adopted from \citealt{hu15}) and Kepler-41b (right; figure adopted from \citealt{shporer15}). The $y$-axes show the flux from the planet, $F_p$, relative to the flux from the host star, $F_s$, in parts per million (ppm), hence it is by definition zero during secondary eclipse when the planet is fully occulted by the star. The overplotted solid line is the atmospheric model, described in \cite{hu15}. The model does not include the transit and the transit data is not plotted. The phase shift between maximum flux and superior conjunction clearly seen in these, and other, optical phase curves imply an inhomogeneous cloud coverage in the planets' atmosphere where reflective condensation clouds are located westward of the substellar point, and the atmosphere east of the substellar point is clear \citep{demory13, hu15, shporer15, parmentier16}.
}
\end{center}
\end{figure*}

For three of these planets, Kepler-7b, Kepler-12b, and Kepler-41b, the optical phase curves are dominated by the atmospheric component, while the two gravitational components' amplitude are an order of magnitude or more smaller and at the data noise level. Therefore these phase curves allow a direct view of the atmospheric phase component shape, independent of any approximations done in determining the shape of the other components (see also \secr{ellip} and \secr{hotstarstide}). The three phase curves show a visually apparent phase shift \citep{demory13, hu15, shporer15} and two of them are shown in \figr{phaseshift}.

\begin{figure}
\hspace{-6mm}
\includegraphics[scale=0.28]{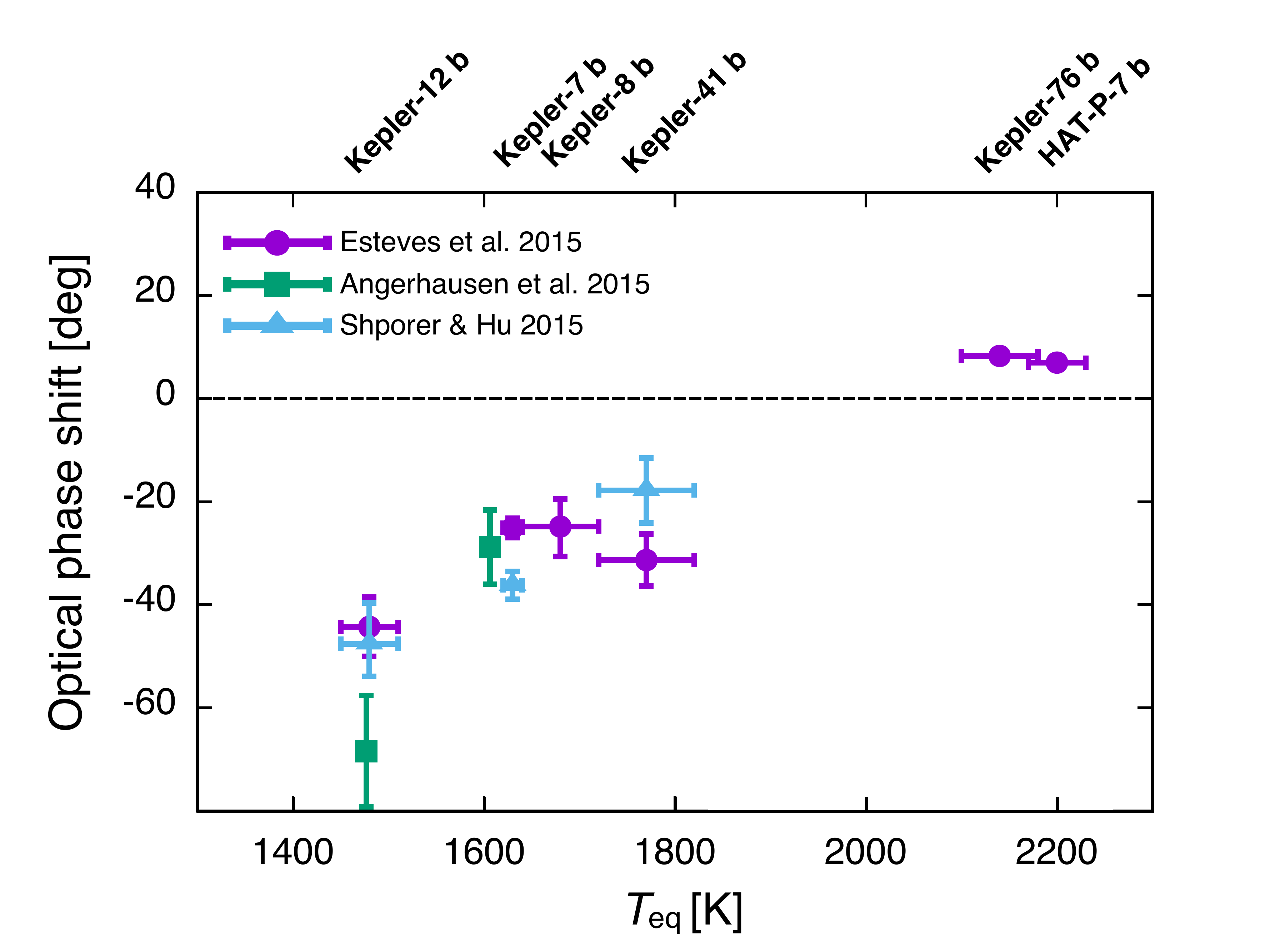}
\caption{
\label{fig:offset}
Optical phase shift between the atmospheric phase component maximum flux and superior conjunction (secondary eclipse), in degrees, against planetary atmosphere equilibrium temperature calculated assuming all incident flux is thermalized (zero Bond albedo). The figure shows that results from three separate studies (see legend) point to a correlation between the two parameters. Figure adopted from \cite{parmentier16}.
}
\end{figure}

A phase shift was measured for a total of 6 planets \citep{angerhausen15, esteves15, shporer15}, and they show a correlation between the phase shifts direction, or sign, with the planet's equilibrium temperature \teq\ \citep{esteves15, parmentier16}. That correlation is presented in \figr{offset} which was adopted from \citet{parmentier16}. Hotter planets show maximum flux at an earlier phase than superior conjunction, as seen in IR phase curves, and cooler planets show maximum flux at a later phase (including the two phase curves shown in \figr{phaseshift}). The phase shift of hot planets is explained as thermal emission from a hot spot eastward of the substellar point, consistent with the theory of phase shifts seen in IR phase curves. The phase shift of the cooler planets is explained by increased reflectivity in the area westward of the substellar point (a reflective spot, or optical bright spot), rotating into the observer's view \emph{later} than superior conjunction. The increased reflectively follows from cloud condensation in the cooler area while the atmosphere is clear in the hotter area \citep[e.g.,][]{demory13, hu15, shporer15}.
 
The clouds chemical composition is difficult to determine uniquely with the \ik\ data alone. However, as shown by \citet[][see also \citealt{munoz15} and \citealt{oreshenko16}]{parmentier16} candidate species can be identified based on 3D global circulation planetary atmosphere models, the atmospheric \teq, and the measured optical phase shift. Therefore, broad band optical phase curves can constrain not only the dynamics and structure of the atmosphere but also its composition.

\subsection{The mass discrepancy}
\label{sec:massdiscrep}

The amplitudes of the two gravitational phase components, beaming and ellipsoidal distortion (see \eqr{abeam2} and \eqr{ellip}, respectively), can each be used to estimate the companion minimum mass when fitted as independent parameters. For transiting systems where $\sin i \approx 1$ these two estimates should agree with each other. However, for a growing list of systems it has been shown that the two mass estimates do not agree. We refer to this disagreement as \emph{the mass discrepancy}. These systems include transiting hot Jupiters: 
TrES-2b \citep{barclay12}, 
HAT-P-7b \citep{esteves13}, 
Kepler-13Ab \citep{shporer11, mazeh12, esteves13, shporer14}, 
Kepler-76b \citep{faigler13}, 
and stellar eclipsing binaries: 
KOI-74 \citep{rowe10, vankerkwijk10, ehrenreich11, bloemen12}, 
KIC 10657664 \citep{carter11}, 
KIC 9164561 \citep{rappaport15}.
For these systems the companion mass estimated based on one phase component amplitude is \emph{not} systematically larger or smaller than the mass estimated based on the other phase component amplitude. For some of the systems above the companion mass was measured using RVs. For those, the companion mass derived using RVs does not systematically agree with the companion mass estimated based on one of the phase components.

A possible reason for such a mass discrepancy is an incomplete understanding of the host star since \Abea\ and \Aell\ depend differently on the stellar mass and radius. However, this is not a likely possibility as the stellar parameters were estimated using spectroscopy for most systems mentioned above, and using asteroseismology for TrES-2 \citep{barclay12} and HAT-P-7  \citep{vaneylen12}. For Kepler-13A the improved stellar parameters derived through spectroscopy have indeed brought the two mass estimates closer but they are still statistically discrepant \citep{shporer14}.

The phase shift of the atmospheric phase component described in \secr{atm} for optical phase curves can cause a mass estimate discrepancy. As the beaming and atmospheric phase components are the sine and cosine components of the same period component, a phase shift in one component is degenerate with an amplitude shift in both \citep{esteves15, shporer15, vonparis16}. In this scenario the companion mass estimated using the photometric beaming amplitude will be biased. This matches the findings for some of the star-planet systems mentioned above, namely HAT-P-7b \citep{esteves13}, TrES-2 \citep{barclay12}, and Kepler-76b \citep{faigler13}, where the companion mass was measured using RVs and was found to agree with the mass estimated based on the ellipsoidal photometric amplitude while it does not agree with the mass estimated based on the beaming photometric amplitude. 

\subsubsection{Tidal distortion of non-convective (hot) stars}
\label{sec:hotstarstide}

A phase shift in the atmospheric phase component does not resolve the mass discrepancy for all systems where it was identified (listed above). KOI-74 \citep{rowe10, vankerkwijk10, ehrenreich11, bloemen12} is an eclipsing stellar binary composed of an A-type star and a low-mass hot white dwarf (WD). \cite{bloemen12} showed that the RV-measured mass ratio is consistent with the estimate based on the beaming photometric amplitude, and inconsistent with (and twice as large as) the mass ratio derived from the ellipsoidal amplitude. As discussed by \cite{vankerkwijk10} this is likely to be attributed to the nature of the primary star. Specifically, the A-type star fast rotation and lack of a convective zone mean that the equilibrium-tide approximation --- assuming that vertically displaced fluid at the stellar surface reaches equilibrium --- is not valid. In a study of the ellipsoidal distortion photometric amplitude of stars across a wide mass range \cite{pfahl08} showed that for hot stars, with $M_1 > 1.4\ \msun$, the amplitude is expected to be an order of magnitude or more larger than described in \eqr{ellip}. Still, this does not explain why for KOI-74 the ellipsoidal distortion amplitude predicts a mass ratio that is \emph{smaller} than the RV-measured mass ratio and not larger \citep{bloemen12}. While this issue is not resolved yet, the findings of \cite{bloemen12} for KOI-74 and of \cite{pfahl08} for hot stars serve as a warning against using ellipsoidal distortion amplitudes of hot stars to estimate mass ratios in binary and star-planet systems.




It is interesting to point out that both KIC 10657664 \citep{carter11} and KIC 9164561 \citep{rappaport15}, two of the eclipsing stellar binaries showing the mass discrepancy, are also composed of an A-type star primary and a hot low-mass WD secondary (although in these two systems the WD is younger and still contracting). While these common properties can help identify the reason for the mass discrepancy, this notion is distorted by the fact that for KIC 9164561 the RV-measured mass ratio is consistent with the ellipsoidal amplitude and not with the beaming amplitude.

In addition, Kepler-13A has an A-type star primary and the orbit was not measured with RVs (at least not yet). Therefore, adhering to the warning above we should consider a scenario where the ellipsoidal amplitude does not give the correct planet mass (or mass ratio). On the other hand, following \secr{atm} it is also possible that the beaming amplitude is biased by a phase shift in the atmospheric phase component induced by a hot spot. However, Kepler-13Ab is one of the hottest known gas giant planets \citep{shporer14} and such planets have been observed to have inefficient heat distribution from the day-side to the night-side hemispheres \citep{cowan11, perez13, heng15, schwartz15, komacek16, komacek17}, making a hot spot scenario unlikely. 

Better understanding of the ellipsoidal photometric amplitude of hot stars can come from studying the phase curves of a large sample of short period binaries with an A-type star primary where high-quality space-based time series photometry is available. In those systems the mass ratio can be measured using the beaming amplitude, and/or if possible using RVs. Given the expected RV amplitudes a precision of a few \kms\ is sufficient, which can be achieved with existing facilities and techniques that are already in use or for example the technique of \cite{becker15} that is oriented towards measuring RVs of fast rotating stars. Therefore, phase curves of stellar binaries can be used as a tool in studying the tidal interaction of hot stars.

\section{Future prospects}
\label{sec:future}

\subsection{Future space-based photometric monitoring surveys}
\label{sec:futuresurveys}

The study of orbital phase curves, as presented here, requires high-quality space-based time series photometry, typically provided by transiting planet surveys. While a large amount of data was already provided by such surveys (CoRoT, \ik, K2), more data is expected to be delivered in the future by the NASA TESS mission \citep{ricker14, sullivan15}, to be launched in 2018, and the ESA PLATO mission \citep{rauer14}, to be launched in the mid 2020's. Analysis of future data will benefit from experience and knowledge gained through the work done with currently available data.

\begin{figure}
\includegraphics[scale=0.47]{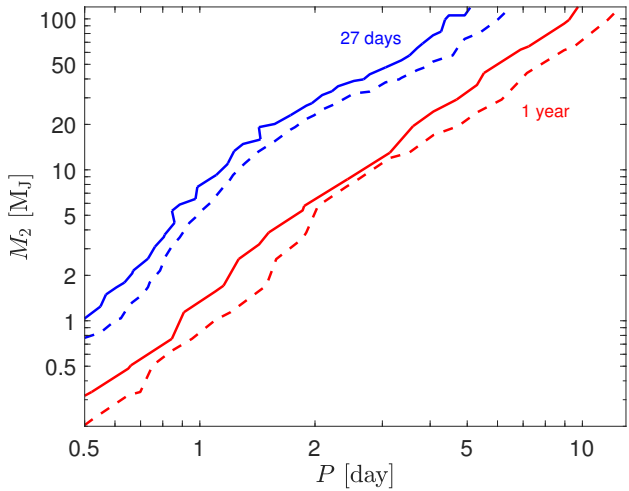}
\caption{
\label{fig:tesssim}
A simple simulation of TESS data sensitivity to orbital phase curves as a function of orbital period ($x$-axis, in days) and companion mass ($y$-axis, in Jupiter mass). The simulation assumes a host star identical to the Sun, a companion with a Jupiter radius, and an orbital inclination of $i$~=~60~deg. The simulation uses half an hour exposures, as expected for TESS full frame images \citep{sullivan15}. The simulation uses a 2D grid in the $M_2-P$ parameter space where in each grid point a phase curve signal is injected to data with the expected TESS noise level for an $I$~=~10~mag star \citep{sullivan15}. The data is generated 100 times and searched for a phase curve signal. The figure shows contours where 90 (dashed lines) and 99 (solid lines) of the simulations resulted in recovering the injected signal. The blue lines present the results for data spanning 27.4 days (a single TESS observation sector), to be obtained by TESS for all bright stars throughout the entire sky, and the red lines present results for 1 year of data (13 TESS observation sectors), to be obtained by TESS for stars close to the celestial poles \citep{sullivan15}. 
}
\end{figure}

\figr{tesssim} shows a simple simulation of the sensitivity of TESS data to non-transiting low-mass companions. We used a 2D grid in the $M_2-P$ parameter space where in each grid position a phase curve signal was injected to random data with the expected TESS noise level for an $I$~=~10~mag star \citep{sullivan15}. The random data was generated 100 times and searched for a phase curve signal. A phase curve was considered as detected when the strongest periodogram peak was at the injected period, and the second strongest peak (ignoring harmonics of the injected period) was less than half the strength of the strongest peak. The simulation assumes the host star is identical to the Sun, the companion has a Jupiter radius, and the orbital inclination is $i$~=~60~deg. The simulation uses half an hour exposures, the expected exposure time of the TESS full frame images \citep{sullivan15}. This temporal resolution is sufficient for detecting the phase curves sinusoidal modulation, since for normal stars the orbital periods are at least an order of magnitude longer. The figure shows contours where 90 (dashed lines) and 99 (solid lines) of the simulations resulted in recovering the injected signal. The blue lines show the results for 27.4~days of data, corresponding to the duration of a single TESS observation sector and which will be obtained for all bright stars throughout the entire sky. The red lines show the results for 1~year of data (356.2~days, equal to 13 TESS observation sectors), that will be obtained by TESS for stars close to the celestial poles \citep{sullivan15}. 

The simulation results show that TESS data will be sensitive to massive brown dwarfs at short periods. We note that the simulation presented in \figr{tesssim} is simple and does not account for cases where the star shows stellar variability that is stronger than the orbital phase curve (generating a periodogram peak that is stronger than that of the orbital signal), and false positives where the stellar variability mimics an orbital phase curve signal. On the other hand, the follow-up of TESS candidates will benefit from the availability of Gaia data  \citep{perryman01, gaia16}, including parallax, spectroscopy, and low-precision RVs, that will give a better understanding of the host stars and at least in some cases identify the large RV orbital modulation of stellar binaries. In addition, TESS targets will be brighter than \ik\ targets, with a magnitude range of about 8~--~12~mag, so follow-up observations can be done with smaller telescopes, for example the Las Cumbres Observatory network (LCO; \citealt{brown13}), and the Network of Robotic Echelle Spectrographs (NRES; \citealt{siverd16}) that will mounted on LCO's 1~m telescopes during 2017.

Another use of TESS phase curves will be to identify a false positive scenario where a short period planetary transit-like signal is caused by a stellar companion whose mass is measured though the phase curve shape.

The ESA PLATO Mission, scheduled to be launched in the mid 2020's, is an ambitious mission that is planned to have a wide field of view, of 2232~deg$^2$, and high temporal resolution, of 25 and 2.5~seconds \citep{rauer14}. The planned observing strategy includes two fields with long duration monitoring of 2-3 years each (i.e.~\ik-like), and a few fields with short duration monitoring of 2--5~months each (i.e.~K2-like). PLATO will observe bright stars, similarly to TESS, and the high temporal resolution will be used for measuring stellar properties through asteroseismology, including mass, radius, and age. Therefore, PLATO data quality and volume are expected to be beyond that of \ik, while the ground-based follow-up of candidates (transiting planet candidates and non-transiting candidates detected through orbital phase modulations) will be more efficient and can be done with small telescopes.

\subsection{Multi band phase curves}
\label{sec:multi}

Almost all currently available phase curve data was measured in a single wide band. However, the phase curve shape is wavelength dependent. The relative contribution of thermal emission and reflected light to the atmospheric phase component is wavelength dependent. 
In addition, the beaming and ellipsoidal phase components also depend on wavelength through the \albea\ and \alell\ coefficients (see \eqr{albeanu} and \eqr{alell}).

For atmospheres of cool objects (low-mass stars, brown dwarfs, and planets) the wavelength dependent phase curve information (e.g.~phase curve amplitude) is sensitive to the atmosphere's vertical profile, in addition to the longitudinal information the phase curve holds. Therefore the phase curve shape as a function of wavelength can provide information about the atmospheric structure, dynamics, and composition \citep[e.g.,][]{stevenson14, parmentier16, oreshenko16, wong16}. That information cannot be obtained from, and is therefore complementary to, transmission and emission spectroscopy.

For a few hot Jupiter planets the phase curve was measured in more than a single IR band with \is\ \citep{knutson12, cowan12, maxted13, lewis13, wong15, wong16}. The shape of a thermal (IR) single-band phase curve is determined by the longitudinal dependence of the atmospheric temperature and composition. Separating the two requires additional information such as multi-band phase curves. In a few of the multi-band IR phase curves obtained so far (HAT-P-7b, WASP-14b, WASP-19b; \citealt{wong15, wong16}) the emission from the night-side hemisphere is significantly smaller than theoretical predictions in the 3.6~\mic\ band while significantly larger than the same predictions in the 4.5~\mic\ band. Such increased opacity in 3.6~\mic\ and decreased opacity in 4.5~\mic\ can be caused by increased C/O ratio in the planet's atmosphere compared to the Solar value of $\approx 0.55$ \citep{moses13}. A supersolar C/O ratio leads to an increased amount of CH$_4$ and decreased amount of CO in the planet's atmosphere \citep{moses13}. CH$_4$ has a vibrational band at 3.6~\mic, explaining the increased opacity, while CO has a vibrational band at 4.5~\mic, explaining the decreased opacity. A supersolar C/O ratio can also explain the low water abundance identified for some hot Jupiters, as described by \cite{moses13}. Future observations will test the increased C/O ratio hypothesis.

So far the phase curve of only a single planet, HAT-P-7b, was measured in both the optical (\ik) and IR (\is\ 3.6~\mic\ and 4.5~\mic\ bands; \citealt{wong16}). The \ik\ data shows a phase curve maximum earlier than the secondary eclipse, indicating an eastward shifted hotspot and thermal emission dominating reflected light in the optical phase curve of this hot Jupiter \citep{esteves15, faigler15}. The \is\ data is less sensitive to such an offset and no statistically significant offset is measured in both IR bands \citep{wong16}.

A first (and so far only) spectroscopic phase curve, where a spectrum was measured throughout the entire orbital phase, was obtained by \cite{stevenson14} for WASP-43b at 1.1~--~1.7~\mic\ using HST WFC3. That work exemplified the potential of spectroscopic phase curves in characterizing the planet's atmosphere. By measuring the wavelength dependency of the phase curve amplitude and the shift between phase of maximum light and secondary eclipse \cite{stevenson14} were able to investigate WASP-43b atmospheric pressure-temperature profile, global energy budget, and chemical composition. They concluded that it has a relatively low Bond albedo of 0.18$^{+0.07}_{-0.12}$, similar to estimates for other hot Jupiters, a highly inefficient heat redistribution from the day to night hemispheres (day-night luminosity difference $> 20$ at \sig{1}), and that water vapor is present in the pressure regions probed by their data.

Future exoplanet characterization missions, like the proposed FINESSE, EChO, and ARIEL missions \citep{deroo12, tinetti15, tinetti16}, will be able to measure spectroscopic phase curves with a high spectral resolution across a wide wavelength range from the optical to the IR. Unlike current space-based facilities (HST, \is) such missions will be specifically designed for this task and will have the telescope time to measure the spectroscopic phase curve for a sample of planets. Phase curve observations are obviously much more time consuming than transmission and emission spectroscopy hence targets will need to be selected carefully. Although, in some circumstances, such as atmospheres with a high mean molecular weight \citep{koll16}, the transmission spectroscopy signal is expected to be small compared to the phase curve signal, requiring repeated transit observations and using a similar amount of time as for measuring the phase curve \citep{koll16}.

A possible way to reduce the amount of time needed for measuring a phase curve is to observe the phase curve intermittently and not continuously, as explored for \is\ by \citet[][see also \citealt{cowan07, crossfield10}]{krick16}. Future general-purpose space telescopes like the {\it James Webb Space Telescope} ({\it JWST}; \citealt{gardner06}; scheduled to be launched in October 2018) will also be capable of measuring a spectroscopic phase curve \citep[e.g.,][]{beichman14, cowan15}, by combing data from different instruments to get a wide wavelength coverage. Although it is not likely that time on {\it JWST} will be awarded to observe more than very few phase curves, {\it JWST} will be launched in the near future, years before any exoplanet characterization mission is launched. Hence {\it JWST} will have the opportunity to obtain the best spectroscopic phase curves for the next few years if not longer. For example, it will be able to test the enhanced C/O ratio hypothesis mentioned above.

Interestingly, some wavelength information can be obtained by comparing \ik\ and TESS phase curves of the same objects, since the transmission curves of the two instruments only partially overlaps \citep{placek16}. While TESS goes deeper into the red end of the visible wavelength range \ik\ goes deeper in the blue end.

\section{Summary}
\label{sec:sum}

The field of (continuous) optical time series photometry is now at its golden age, manifested by the continuum of space-based photometric surveys, spanning from the concluded CoRoT and \ik\ missions, through the ongoing K2 mission, to the near-future TESS mission and far-future PLATO mission. The study of orbital phase curves takes advantage of the high data quality and the scientific opportunities they unveil.

As shown here orbital phase curves in the optical can be a used as an astrophysical tool, especially in areas where the traditional tools and approaches have proved inefficient, such as the study of short-period low-mass companions to hot stars and the study of short-period brown dwarf companions around stars across the main sequence. Those phase curve-based methods are now being developed and perfected. 

However, at the same time these applications have proven that at least in some cases the understanding of the phase curve shape is incomplete. While that warrants caution it is also an opportunity to learn new science. Therefore going forward orbital optical phase curves should be considered as a useful astrophysical tool on one hand, and on the other hand a subject of detailed study in their own right.

Current and future surveys are expected to deliver more high quality broad band data and spectroscopic data that will lead the field of orbital phase curves from its current infancy to adulthood.

\acknowledgments

I am grateful to the anonymous referee for their thorough reading of the manuscript and useful comments that helped improve it.
While preparing this manuscript I have benefitted from discussions and feedback from numerous colleagues, including: Jim Fuller, Renyu Hu, Heather Knutson, Daniel Koll, Yossi Schwartzvald, Lev Tal-Or, and Ian Wong. 
I would like to warmly thank Steven Bloemen, Simchon Faigler, and Josh Winn for providing comments on earlier versions of the manuscript.
This work was performed in part at the Jet Propulsion Laboratory, under contract with the California Institute of Technology (Caltech) funded by NASA through the Sagan Fellowship Program executed by the NASA Exoplanet Science Institute.
This research has made use of NASA's Astrophysics Data System Service.

\bibliography{bib}

\end{document}